\documentclass[twocolumn,aps,amsmath,amssymb,floatfix,superscriptaddress,prb,longbibliography]{revtex4-2}

\usepackage{graphicx}
\usepackage{dcolumn}
\usepackage{bm}
\usepackage{xspace}
\usepackage[colorlinks]{hyperref}
\usepackage{color}
\usepackage[flushleft]{threeparttable}
\def\MPS{MnPSe$_3$\xspace}
\def\new{\textcolor{black}}
\def\rnew{\textcolor{black}}

\begin{document}
	
	
	\title{Spin and lattice dynamics of a van der Waals antiferromagnet MnPSe$_3$}
	\author{Junbo Liao}
	\author{Zhentao~Huang}
	\author{Yanyan Shangguan}
	\author{Bo~Zhang}
	\author{Shufan~Cheng}
	\author{Hao~Xu}
	\affiliation{National Laboratory of Solid State Microstructures and Department of Physics, Nanjing University, Nanjing 210093, China}
	\author{Ryoichi Kajimoto}
	\affiliation{J-PARC Center, Japan Atomic Energy Agency (JAEA), Tokai, Ibaraki 319-1195, Japan}
	\author{Kazuya Kamazawa}
	\affiliation{Neutron Science and Technology Center, Comprehensive Research Organization for Science and Society (CROSS), Tokai 319-1106, Ibaraki, Japan}
	\author{Song~Bao}
\email{songbao@nju.edu.cn}
\affiliation{National Laboratory of Solid State Microstructures and Department of Physics, Nanjing University, Nanjing 210093, China}
	\author{Jinsheng~Wen}
	\email{jwen@nju.edu.cn}
	\affiliation{National Laboratory of Solid State Microstructures and Department of Physics, Nanjing University, Nanjing 210093, China}
	\affiliation{Collaborative Innovation Center of Advanced Microstructures, Nanjing University, Nanjing 210093, China}

	
\begin{abstract}
Antiferromagnetic van der Waals family $\rm \textit{M}P\textit{X}_{3}\ (M=Fe,\ Mn,\ Co,\text{ and}\ Ni; X=S\text{ and}\ Se)$ have attracted significant research attention due to the possibility of realizing long-range magnetic order down to the monolayer limit. Here, we perform inelastic neutron scattering measurements on single crystal samples of \MPS, a member of the $\rm \textit{M}P\textit{X}_{3}$ family, to study the spin dynamics and determine the effective spin model. The excited magnon bands are well characterized by \new{a spin model, which includes a Heisenberg term with three intraplane exchange parameters ($J_{1}=-0.73$~meV, $J_{2}=-0.014$~meV, $J_{3}=-0.43$~meV) and one interplane parameter ($J_{c}=-0.054$~meV), and an easy-plane single-ion anisotropy term ($D=-0.035$~meV).} Additionally, we observe the intersection of the magnon and phonon bands but no anomalous spectral features induced by the formation of magnon-phonon hybrid excitations at the intersecting region. We discuss possible reasons for the absence of such hybrid excitations in \MPS.

\end{abstract}

\maketitle

\section{Introduction}
		
\begin{figure*}[htb]
\centerline{\includegraphics[width=7.in]{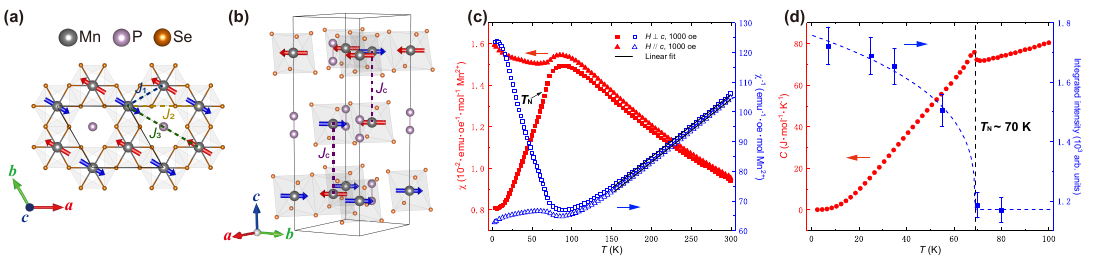}}
\caption{(a) Top view of the hexagonal structure of \MPS in the \it{a-b} \rm plane.  Arrows indicate the magnetic moments on Mn atoms. Dashed lines indicate the paths for the magnetic exchange interactions. (b) Schematic structures of the primitive cell of \MPS. (c) Temperature dependence of the magnetic susceptibility $\chi$ (filled symbols, left axis) and the inverse magnetic susceptibility $\chi^{-1}$ (open symbols, right axis) with field applied parallel and perpendicular to the $c$-axis. Solid lines and the arrow denote the Curie-Weiss fit of $\chi^{-1}$ and the N\'eel transition temperature, respectively. (d) Temperature dependence of the specific heat $C$ (circles, left axis) and the integrated intensity of $(-1,0,\,1)$ Bragg peak (squares, right axis). Error bars represent one standard deviation. The vertical dashed line indicates the N\'eel transition temperature. The dashed curve through the integrated intensities is a guide to the eyes.
\label{fig1}}
\end{figure*}
	
The successful exfoliation of monolayer graphene has ushered in a new era of the research in van der Waals (vdW) materials \cite{Graphene}. Over the past two decades, extensive research on atomic-thin vdW materials has led to the discovery of a wealth of fascinating physical phenomena, as well as tuning and controlling methods \cite{RevModPhys.83.407, liu2016van, RevModPhys.90.041002, RevModPhys.92.021003}. In recent years, the realization of intrinsic two-dimensional (2D) magnetism has established magnetic vdW materials as a promising platform for both application and fundamental research \cite{huang2017layer, gong2017discovery, Intrinsic2DXY}. According to Mermin-Wagner theorem, enhanced fluctuations will prevent the formation of long-range magnetic order in an isotropic low-dimensional magnetic system ($d<3$) \cite{PhysRevLett.17.1133}. Therefore, the key to achieving intrinsic 2D magnetism lies in the exfoliation of bulk magnetic vdW materials with magnetic anisotropy down to the monolayer limit \cite{huang2017layer, gong2017discovery, Intrinsic2DXY}. Presently, several classes of magnetic vdW materials have demonstrated intrinsic 2D magnetism, including vdW ferromagnets such as $\rm Cr_{2}Ge_{2}Te_{6}$~\cite{gong2017discovery}, $\rm Fe_{3}GeTe_{2}$~\cite{deng2018gate,PhysRevX.12.011022} and $\rm Cr\textit{X}_{3}\ (X=Cl,\ Br,\ I)$~\cite{huang2017layer, Intrinsic2DXY}, and vdW antiferromagnets like $\rm \textit{M}P\textit{X}_{3}\ (M=Fe,\ Mn,\ Co,\ Ni; X=S\ Se)$~\cite{lee2016ising, ni2021imaging}.

As an early-discovered antiferromagntic vdW family~\cite{klingen1968hexathio, Prep,WIEDENMANN19811067,NeutDi}, $\rm \textit{M}P\textit{X}_{3}\ (M=Fe,\ Mn,\ Co,\ Ni; X=S\, Se)$ have been extensively studied. Compounds in this family share the same hexagonal monolayer structure [Fig.~\ref{fig1}(a)]. Each $\rm [P_{2}X_{6}]^{4-}$ unit is located in the center of nearby six $\rm M^{2+}$ ions. These $\rm M^{2+}$ ions connected to X atoms form a honeycomb structure, and the atomic layers are weakly coupled through the vdW force. Although these compounds all exhibit an antiferromagnetic order, the magnetic ground state and effective spin model vary significantly when varying the transition metal ions $\rm M^{2+}$. For example, $\rm MnPS_{3}$ exhibits a N\'eel-type spin order and well fits a Heisenberg model~\cite{WIEDENMANN19811067, NeutDi, MagneticPro, PhysRevB.46.5425, A_R_Wildes_1994, A_R_Wildes_1998, PhysRevB.74.094422}. On the other hand, $\rm FePS_{3}$ and $\rm FePSe_{3}$ exhibit a zigzag order and fit an Ising model~\cite{lee2016ising, WIEDENMANN19811067, NeutDi, PhysRevB.76.134402, Wildes_2012, PhysRevB.94.214407,chen2024thermal}. $\rm CoPS_{3}$ and $\rm NiPS_{3}$ also display a zigzag order but are better described by an XY-like model~\cite{PhysRevB.46.5425, PhysRevB.107.054438, PhysRevB.102.184429,PhysRevB.107.054438, scheie2023spin,  PhysRevB.92.224408}. This antiferromagnetic vdW class also provides a platform for tuning the 2D magnetism through doping~\cite{PhysRevMaterials.4.034411,PhysRevMaterials.5.073401}, pressure~\cite{PhysRevX.11.011024,ma2021dimensional}, and strain~\cite{ni2021imaging}, rendering these materials promising application potentials. Moreover, research on the atomic-thin sample reveals exotic physics, such as novel exciton behaviors in $\rm NiPS_{3}$~\cite{kang2020coherent, hwangbo2021highly} and magnon-phonon hybrid excitations, aka magnon polarons in $\rm FePS_{3}$~\cite{PhysRevLett.127.097401, MagnetoRamanStu} and $\rm FePSe_{3}$~\cite{cui2023chirality}.

\MPS is a member of the $\rm \textit{M}P\textit{X}_{3}$ class. It exhibits a N\'eel-type antiferromagnetic order with a transition temperature $T_{\rm N}\backsimeq74 \pm 2$~K~\cite{WIEDENMANN19811067}. Neutron diffraction studies have confirmed a $k=(0, 0, 0)$ propagation vector, with the spin moments lying within the \textit{a-b} plane when the compound is in its antiferromagnetic phase~\cite{LEFLEM1982455,PhysRevMaterials.4.034411}. Although the spin ground state of \MPS has been determined, its effective spin model remains unclear. While some studies suggest an XY model~\cite{PJeevanandam_1999,ni2021imaging}, an inelastic neutron scattering (INS) investigation on powder samples leans towards a Heisenberg model~\cite{PhysRevB.103.024414}. Due to the lack of quantitative and detailed information on the momentum dependence of the spin dynamics, it is challenging to determine the effective spin model based on the results from powder samples. Therefore, a detailed INS research on the single crystal samples is desired to determine the accurate effective spin model of \MPS. In addition, a Raman spectroscopy research reported the observation of the coupling between magnons and the optical phonons in \MPS, indicating the presence of magnon polarons~\cite{mai2021magnon}. Since neutron spectroscopy can probe both magnons and phonons in the momentum-energy space~\cite{squires1996introduction, shirane2002neutron}, it is  an ideal method to detect the possible magnon polarons.

In this work, we carry out INS experiment on single crystal samples of \MPS. We obtain clear magnon bands which exhibit a quasi-2D nature. \new{We find that an effective spin model with three intraplane exchange parameters ($J_{1}=-0.73$~meV, $J_{2}=0.014$~meV, $J_{3}=-0.43$~meV), one interplane parameter ($J_{c}=-0.054$~meV), and one easy-plane single-ion anisotropy term ($D=-0.035$~meV) well describes the spin dynamics.} Additionally, we find that the quasi-2D magnon bands intersect with a phonon band in the vicinity of (0,\,0,\,6), but no evidence of magnon-phonon coupling and the subsequent formation of magnon polarons has been detected. We discuss possible reasons for why these are absent in \MPS.
	
\section{Experimental Details}
	
The crystal structure of \MPS belongs to the space group $R\overline{3}$ (No.~148), consisted of three hexagonal atomic layers arranged in an ABC-type stacking within a unit cell, as shown in Figs.~\ref{fig1}(a) and~\ref{fig1}(b). The lattice parameters are $a=b=6.387$~\AA~ and $c=19.996$~\AA~\cite{WIEDENMANN19811067}. Single crystals of \MPS were grown by the chemical vapor transport method~\cite{PJeevanandam_1999, doi:10.1021/acsnano.5b05927}. As-grown \MPS crystals appeared as thin wine-red flakes and were easy to bend and cut. Typical size of the single crystals was about 5 $\times$ 5 $\times$ 0.1 $\rm mm^{3}$. Magnetization and specific heat measurements were conducted in a Physical Property Measurement System (PPMS-9T) from Quantum Design.
	
For the INS experiments, we glued 140 pieces of single crystals onto aluminum plates using a trademarked fluoropolymer CYTOP-M, with a total weight of 1.86~g. The orientations of the crystals were determined using a back-scattering Laue x-ray diffractometer. The single crystals were mounted on the rectangular aluminum plates, with the [100] and [-120] axes aligned along the edge directions of the aluminum plate. The co-aligned sample had a mosaic spread of $2^{\circ}$.
	
The INS experiment was conducted on a time-of-flight spectrometer 4SEASONS at the Materials and Life Science Experimental Facility of the J-PARC Center in Japan~\cite{Kajimo}. We used a primary energy of $E_{\rm i} = 14.00$ meV and a chopper frequency of 300~Hz to conduct the measurements, resulting in an energy resolution of 0.6~meV at the elastic line. Thanks to the multiple-$E_{\rm i}$ mode equipped on 4SEASONS, we also obtained data with other $E_{\rm i}$s of 7.87, 10.29, and 20.19~meV, with the respective energy resolution of 0.3, 0.4, and 0.9~meV at the elastic positions~\cite{doi:10.1143/JPSJ.78.093002}. We set the angle of the incident neutron beam direction parallel to $c$ axis to be zero. Scattering data were collected by rotating the sample about the [-120] direction in a step of $1^{\circ}$ from $-50^{\circ}$ to $180^{\circ}$ at 6~K, and from $0^{\circ}$ to $105^{\circ}$ at 150~K. At both temperatures, we counted 15 minutes for each step. In addition, measurements were also performed at seven other temperatures of 25, 35, 55, 70, 80 and 120~K. At these temperatures, we fixed the angle to be $0^{\circ}$ and counted one hour for this angle. Raw data were reduced and analyzed using Utsusemi and Horace software~\cite{doi:10.7566/JPSJS.82SA.SA031,EWINGS2016132}. The wave vector $\bm Q$ was expressed as $(H,\,K,\,L)$ reciprocal lattice unit (rlu) of $(a^{*},\,b^{*},\,c^{*})=(4\pi/\sqrt{3}a,\,4\pi/\sqrt{3}b,\,2\pi/c)$. \new{We corrected the measured neutron scattering intensity by dividing the square of the magnetic form factor $|F(\bm Q)|^2$ of $\rm Mn^{2+}$ ions. The corrected data were further divided by the Bose factor via $\chi''(\bm{Q},\omega)=(1-e^{-\hbar\omega/k_{\rm B}T})S(\bm{Q},\omega)$ to eliminate the influence of the Bose statistics, where $k_{\rm B}$ is the Boltzmann factor.} Theoretical simulations of the excitation spectra were performed using the SPINW program~\cite{Toth_2015}.
	
\section{Results}
\subsection{Sample characterizations}
We have conducted the magnetic susceptibility measurements on a single crystal of \MPS with a 0.1-T field applied for both in-plane and out-of-plane directions, and the results are shown in Fig.~\ref{fig1}(c). Upon cooling, we observe little difference in the magnetic susceptibility between the two directions in the paramagnetic phase ($T>120~\rm K$), similar to that in MnPS$_3$ as reported in Ref.~\cite{PhysRevB.46.5425} and~\cite{PJeevanandam_1999}, indicating a weak magnetocrystalline anisotropy in \MPS. We have also noted that in Ref.~\cite{PJeevanandam_1999}, a large anisotropy over the entire temperature range was reported for \MPS. We have measured several single crystals we grew, and all of them show similar results to those in Fig.~\ref{fig1}(c). As we will discuss in detail below, this weak anisotropy is consistent with our neutron scattering results. In the vicinity of 100~K, the susceptibility displays a broad peak, which is a typical feature of low-dimensional magnetic systems~\cite{PhysRevB.46.5425}. This peak is ascribed to the establishment of short-range spin-spin correlations among the magnetic moments in the materials~\cite{PhysRevB.46.5425}. At lower temperatures, long-range magnetic order occurs, as evidenced by the kink point around 70~K in the magnetic susceptibility curves [Fig.~\ref{fig1}(c)]. This magnetic phase transition at the $T_{\rm N}\backsimeq70$~K is further supported by the $\lambda$-type anomaly observed in the specific heat measurements and the onset of magnetic scattering intensities at the magnetic Bragg peak (-1,\,0,\,1) as observed in the neutron scattering measurements [Fig.~\ref{fig1}(d)]. Upon further cooling, the in-plane susceptibility dramatically decreases, while the out-of-plane susceptibility slightly increases~[Fig.~\ref{fig1}(c)]. Such behaviors indicate that \MPS establishes a collinear antiferromagnetic order with the magnetic moments lying in the \textit{a-b} plane, which is consistent with previous neutron diffraction results \cite{WIEDENMANN19811067,PhysRevB.103.024414}.
	
To quantitatively understand the magnetic susceptibility results, we fit the data above 150~K using the Curie-Weiss law, $1/\chi=(T+\Theta)/C$, where $C$ and $\Theta$ are the Curie constant and Curie-Weiss temperature, respectively [Fig.~\ref{fig1}(c)]. The extracted Curie constants are 4.72 and 4.75~$\rm emu\ K/mol\ Mn^{2+}$, and the resulting effective magnetic moments of 6.14 and $6.16\ \mu_{\rm B}$ for $H|| c$ and $H\perp c$, respectively. These results indicate a high spin state of $S=5/2$ for Mn$^{2+}$ ions and a Land\'e g factor of 2.08 with a negligible spin-orbit coupling in this material. The fitted Curie-Weiss temperatures are $\Theta_{\parallel}=-192.06~\rm K$ and $\Theta_{\perp}=-203.10~\rm K$, respectively, demonstrating strong antiferromagnetic correlations in \MPS.
	
\begin{figure}[htb]
		\centerline{\includegraphics[width=3.in]{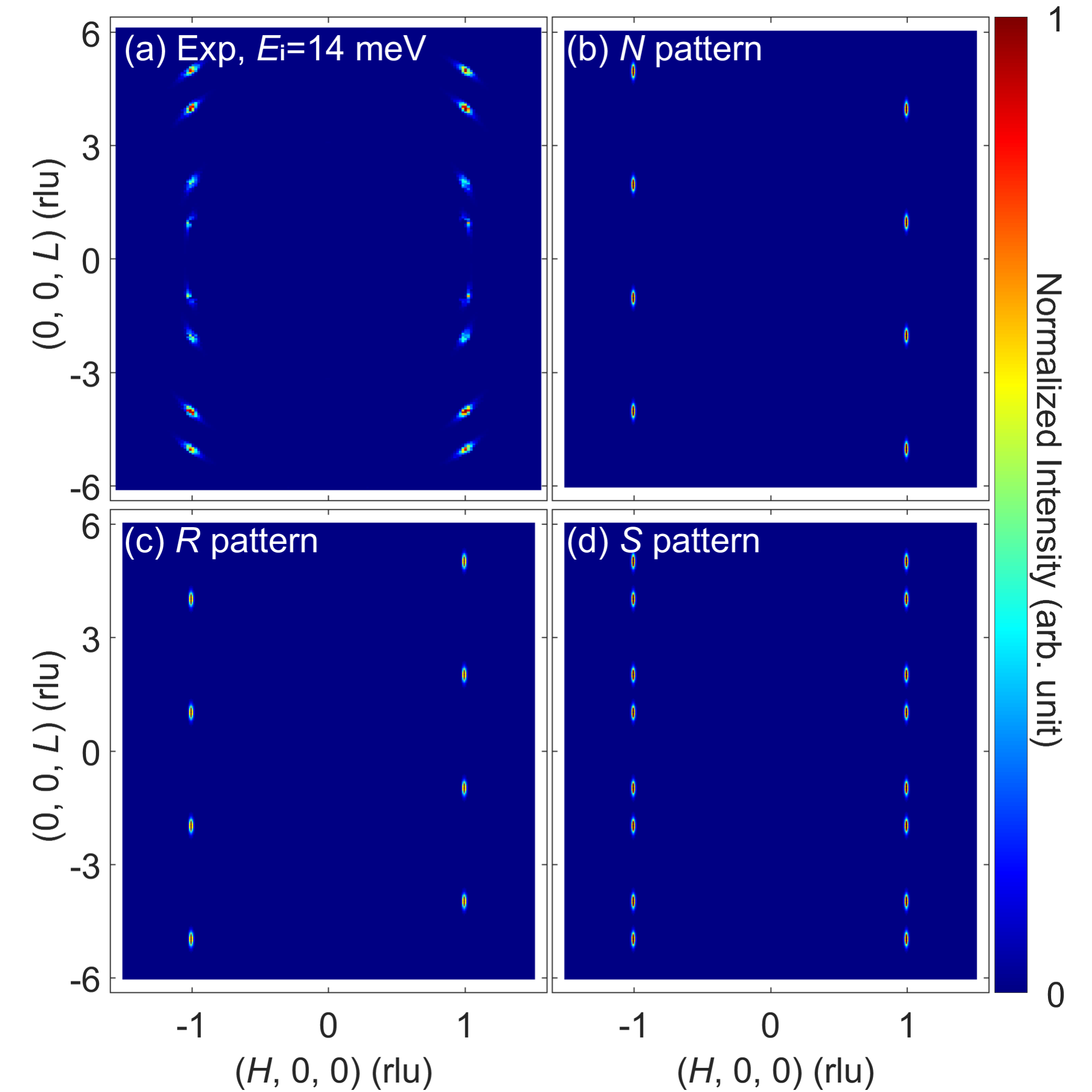}}
		\caption{ (a) Measured magnetic Bragg peaks in the $(H,\,0,\,L)$ plane with the incident energy of $E_{\rm i}=14$~meV for \MPS. The energy is integrated over [-0.1,\,0.1]~meV. The lattice contributions are eliminated by subtracting the data of 150~K. (b) Theoretically calculated magnetic Bragg peaks in the $(H,\,0,\,L)$ plane according to the untwinned case, termed $N$ pattern. (c) Reflection of the $N$ pattern about the $L=0$ line, termed $R$ pattern. (d) Superposition of the patterns in (b) and (c), termed $S$ pattern, representing the magnetic Bragg scatterings from the twinned sample.
			\label{fig2}}
\end{figure}
	
\begin{figure*}[htb]
		\centerline{\includegraphics[width=7.in]{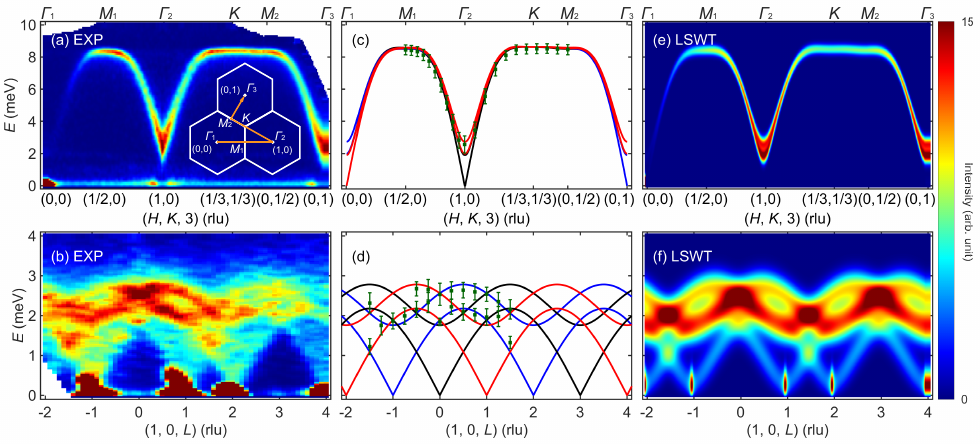}}
		\caption{ \new{INS results of the magnetic excitation spectra at 6~K along the in-plane (a) and out-of-plane (b) directions. The data were obtained with $E_{\rm i}=14$~meV for (a), and $E_{\rm i}=7.87$ meV for (b). In (a), the integration thickness of the other in-plane direction, orthogonal to the high-symmetry path, is chosen to be $\pm$0.05~rlu, and the wave vectors $L$s are integrated over [2.5,\,3.5]~rlu. In (b), the integration range for the two orthogonal in-plane vectors is $\pm$0.05~rlu. (c) and (d), Calculated in-plane and out-of-plane magnon dispersions using the linear spin-wave theory (LSWT), respectively. The data points are extracted from the experimental spectra presented in (a) and (b). (e) and (f), Calculated magnetic excitation spectra, which are superposed with $\pm L$ cases taking into account the twinning effect and instrumental resolutions. The inset in (a) illustrates the high-symmetry paths of the in-plane excitation spectra.}
			\label{fig3}}
\end{figure*}
	
\subsection{Magnon spectra}
\new{Although the crystal structure of \MPS possesses a $\bar{C_3}$ axis, the Laue pattern shows a nearly $C_6$ symmetry, which makes it difficult to distinguish two kinds of sample alignments rotated by 60 degree. Due to the $\bar{C_3}$ axis, these two alignments are equivalently connected by the mirror operation about the $a$-$b$ plane. Therefore, the data we have collected with a nominal ${\bm Q}=(H,\,K,\,L)$ corresponds to a superposition of the actual signals at $(H,\,K,\,L)$ and $(H,\,K,\,-L)$, which we call the ``twinned" effect.} This can be confirmed by comparing the distribution of experimentally measured and theoretically predicted magnetic Bragg peaks in the $(H,\,0,\,L)$ plane. Figure~\ref{fig2}(a) shows the measured magnetic Bragg peaks in the $(H,\,0,\,L)$ plane. Given that the spin moments in the sublattice of each layer align parallel to those in the adjacent layers and the number of layers per unit cell being three, the magnetic Bragg peaks are expected to be located at integer positions of $L$ with a periodicity of 3~rlu along the [001] direction, which is indeed observed in Fig.~\ref{fig2}(a). These peaks are symmetrically distributed about the $L=0$ line. Figure~\ref{fig2}(b) shows the theoretically calculated distribution of the magnetic Bragg peaks without taking into account the twinned effect. Different from the experimental results, the calculated pattern does not show reflection symmetry about the $L=0$ line but instead exhibits centrosymmetry about the origin (0,\,0,\,0). For clarity, we refer to this pattern as the \textit{N} pattern, where \textit{N} stands for ``normal''. In Fig.~\ref{fig2}(c), we present the reflected results of the \textit{N}  pattern about the $L=0$ line and refer to it as the \textit{R} pattern. The comparable intensities observed in the two sets of magnetic Bragg peaks in Fig.~\ref{fig2}(a) suggest that the single crystals contributing to the \textit{N} and \textit{R} patterns are nearly equal. Therefore, we superpose these two patterns in a 1:1 ratio, and term it \textit{S} pattern. The results are shown in Fig.~\ref{fig2}(d). Obviously, the \textit{S} pattern reproduces the experimental results well, supporting the conjecture regarding the sample twinning.

After clarifying the twinned issue of the sample alignment, we now present the spin excitation spectra of \MPS. \new{Figures~\ref{fig3}(a) and~\ref{fig3}(b) show the in-plane and out-of-plane spin excitation spectra, respectively. The paths for the in-plane spin excitation spectra are illustrated in the inset of Fig.~\ref{fig3}(a).} It is found that the well-defined magnon excitations disperse up from the zone center $\Gamma$ point and propagate towards the zone boundary, with the maximum occurring around 8.5~meV, consistent with previous neutron scattering results from a powder sample~\cite{PhysRevB.103.024414}. This profile closely resembles that of its sister compound $\rm MnPS_{3}$, which has an 11.5-meV bandwidth, indicating remarkable intralayer exchange interactions~\cite{A_R_Wildes_1998}. It is noteworthy that the magnons exhibit an onset of excitations around 1.8~meV for $L=3$~rlu~[Fig.~\ref{fig3}(a)]. \new{However, this does not indicate an excitation gap but is attributed to the out-of-plane variations of the magnetic excitations, which mainly modify the bottom of the magnon dispersions. This can be visualized from the excitations along the [1, \,0, \,$L$] direction as plotted in Fig.~\ref{fig3}(b). Note that the out-of-plane spectra show remarkable broadening in momentum, which indicates the weak correlations along $L$ due to the two dimensionality. Similar to the elastic patterns shown in Fig.~\ref{fig2}(a), these excitations show an $L=3$~rlu period. In the range of $L=[-1, 1]$~rlu, two winding bands can be recognized in the energy range of [1.9, 2.8]~meV. As will be confirmed in next section, these two bands actually correspond to $N$ and $R$ bands due to the sample twinning along the $L$ direction. In addition, the spectral weight is continuously distributed towards the magnetic Bragg peaks at (1, \,0, \,$\pm$1+3$n$) rlu. Although the detailed spectra below 0.3~meV are obscured by the strong and elongated signal of the magnetic Bragg peaks, making it difficult to determine whether there is an excitation gap, our data sets an upper bound of the excitation gap of 0.3~meV. Moreover, a recent study employing magneto-spectroscopy reported the gap will be less than 0.1~meV, if it does exist~\cite{jana2023magnon}.}

\begin{table*}[t]
\caption{\label{table1}%
The fitting exchange parameters obtained from the current work on single-crystal samples and reported results from powder samples in Ref.~\cite{PhysRevB.103.024414}.  \rnew{Error bars indicate one standard deviation uncertainty.} The goodness of fit is represented by the value of $\chi ^{2}$, where a value closer to 1 indicates a better fit.}

\begin{ruledtabular}
\begin{tabular}{lcccccccccc}
Paper & Model & \textrm{$J_{1}$ (meV)}&\textrm{$J_{2}$ (meV)}& \textrm{$J_{3}$ (meV)}& \textrm{$J_{c} $(meV)} &$D$ (meV)&$\delta$& $\chi^{2}$\\
\colrule
Current work & Heisenberg + D & \rnew{-0.73~$\pm$~0.03} & \rnew{-0.014~$\pm$~0.005} & \rnew{-0.43~$\pm$~0.04} & \rnew{-0.054~$\pm$~0.005} & \rnew{-0.035~$\pm$~0.004} & $-$ & \rnew{1.5}\\
Current work & Heisenberg & \rnew{-0.74~$\pm$~0.03} & \rnew{-0.017~$\pm$~0.005} & \rnew{-0.41~$\pm$~0.05} & \rnew{-0.075~$\pm$~0.004} & $-$ & $-$ &\rnew{2.4}\\
Ref.~\cite{PhysRevB.103.024414} & Heisenberg & -0.9 & -0.06 & -0.38 & -0.062&$-$& $-$ &\rnew{3.8}\\
Current work & XXZ & -0.73 & -0.014 & -0.43 & -0.054 & $-$ & 0.981 & $-$\\
\end{tabular}
\end{ruledtabular}

\end{table*}

\subsection{Effective spin model}
To capture the observed features of the well-defined magnons and determine the effective spin model, \new{we employ the following spin Hamiltonian to model our data, considering the easy-plane magnetic order: 
	\begin{equation}\label{spinw}
		H=-\sum_{\langle i,j \rangle} J_{i,j}{\bm{S}}_{i}\cdot{\bm{S}}_{j}-\sum_{\langle i \rangle} D_{i}({S}^{z}_{i})^{2}.
	\end{equation}
Here, ${S}_{i}$ is spin operators of a $\rm Mn^{2+}$ ion at the site $\bm{r}_{i}$ with a magnitude of 5/2, $J_{i,j}$ is the exchange interaction connecting two $\rm Mn^{2+}$ ions at sites $\bm{r}_{i}$ and $\bm{r}_{j}$, and ${D}_{i}$ is the single-ion anisotropy of $\rm Mn^{2+}$ ions.} We take three intralayer exchange interactions (nearest-neighbor interaction $J_{1}$, next-nearest-neighbor interaction $J_{2}$ and third-nearest-neighbor interaction $J_{3}$) and one interlayer exchange interaction ($J_{c}$) into consideration in the parameterization. The detailed configuration of these interactions are given in Figs.~\ref{fig1}(a) and~\ref{fig1}(b). We use the SPINW package to model our data with the linear spin-wave theory (LSWT)~\cite{nolting2009quantum,Toth_2015}. In this approach, a standard Holstein-Primakoff transformation is applied to Eq.~\ref{spinw} to get a $12 \times 12$ matrix that characterizes the spin dynamics of \MPS. The dispersion relation $\hbar \omega(\bm Q)$ is obtained by a Bogoliubov transformation on the $12 \times 12$ matrix.

\new{To determine the specific exchange parameters $J_{i,j}$ and single-ion anisotropy ${D}_{i}$}, we firstly make constant-$\bm Q$ cuts through the excitation spectra in Figs.~\ref{fig3}(a)-\ref{fig3}(b) at various momenta and fit these cuts with Gaussian line shapes to extract the magnon energies. By doing so, we can obtain the magnon dispersion. Subsequently, we fit the extracted dispersion relation with a theoretical $\hbar \omega(\bm Q)$ function. \new{In this procedure, we restrict $D_{i}$ to be negative to keep consistent with the easy-plane anisotropy, revealed by the easy-plane magnetic order and negligible excitation gap~\cite{jana2023magnon}.} The refined exchange parameters are presented in the first row of Table.~\ref{table1}, where the additional parameter $\chi ^{2}$ evaluates the goodness of the fits in the SPINW program. \new{Some extracted experimental points and the calculated dispersions with the refined parameters are plotted in Figs.~\ref{fig3}(c) and~\ref{fig3}(d). The calculated dispersions fit the extracted experimental points quite well. With the involved magnetic parameters, the calculations give six magnon bands, consistent with the 6 magnetic ions in a unit cell. These six bands can be divided into three groups, denoted by red, blue, and black colors in Figs.~\ref{fig3}(c) and~\ref{fig3}(d). Each group is consisted of one acoustic and one optical branch. These three dispersion groups have the same shape, but are shifted by $\Delta L=1$~rlu from each other along the [001] direction, which can be visualized from Fig.~\ref{fig3}(d).}

\begin{figure*}[htb]
\centerline{\includegraphics[width=7.in]{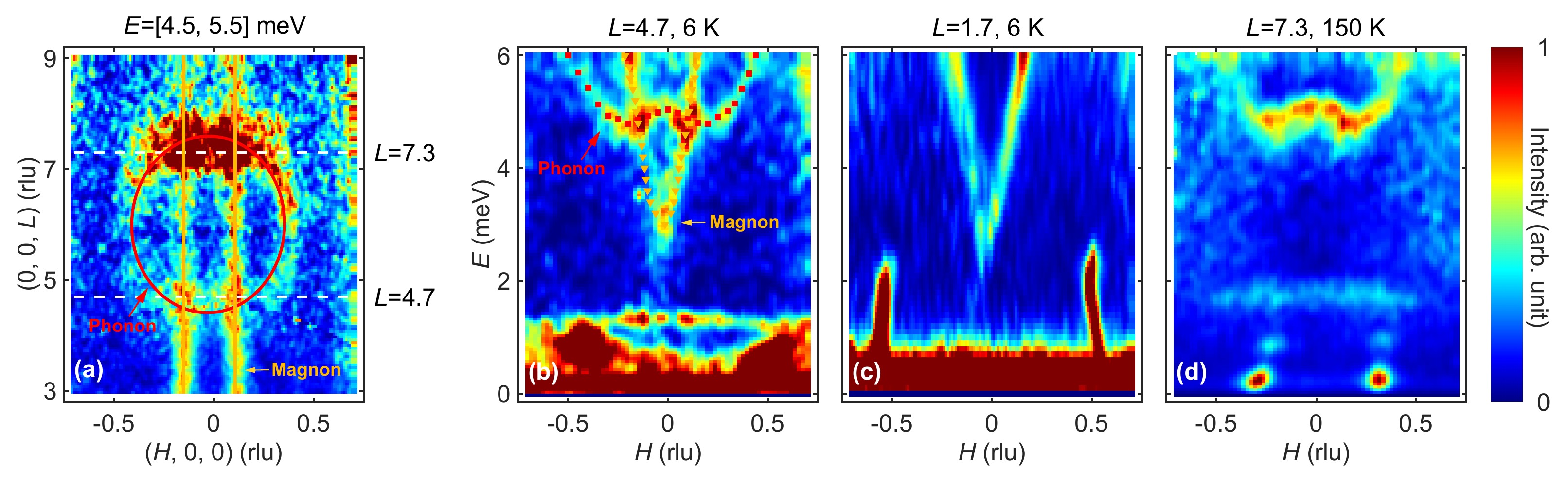}}
\caption{ (a) Constant-$E$ contour at 5~meV in the $(H,\,0,\,L)$ plane, obtained with $E_{i}=10.29$ meV. The integration thickness of the energy is $\pm 0.5$~meV. lines and oval denote magnon excitations and phonon excitaions, respectively. The dashed lines indicate the $L$ positions where the magnon and phonon dispersions intersect with each other. (b) The overlapped magnon and phonon excitations at 6~K presented in the same momentum-energy window with $L=4.7$~meV. Triangles and squares denote magnon and phonon dispersions extracted from (c) and (d), respectively. (c) and (d), The individual magnon and phonon spectra, obtained with $L=1.7$~rlu at 6~K and $L=7.3$~rlu at 150~K, respectively. The integration thicknesses of the [-120] and [001] directions is $\pm 0.1$ and $\pm 0.15$~rlu, respectively.
\label{fig4}}
\end{figure*}

In order to directly compare the intensities with the experimental results, we calculate the dynamical spin-spin correlation function \cite{squires1996introduction, shirane2002neutron},
	\begin{equation}
		S^{\alpha \beta}(\bm{Q},\omega)=\frac{1}{2\pi}\sum_{l}e^{i\bm{Q}\cdot \bm{r}_{l}}\int_{-\infty}^{\infty}\langle S^{\alpha}_{0}(0) S^{\beta}_{l}(t) \rangle e^{-i\omega t} dt.
	\end{equation}
$S^{\alpha \beta}(\bm{Q},\omega)$ is related to the scattering cross section as,
	\begin{equation}
		\frac{d^2\sigma}{d\Omega d\omega}\varpropto |\bm{F}(\bm{Q})|^{2}\sum_{\alpha \beta}(\delta_{\alpha \beta}-\frac{Q_{\alpha}Q_{\beta}}{Q^2})S^{\alpha \beta}(\bm{Q},\omega),
	\end{equation}
which is directly measured in INS experiments. \new{The calculated results of $S^{\alpha \beta}(\bm{Q},\omega)$ show that only the two bands in red group possess significant spectral weight, while the rest two groups are invisible in principle. However, due to the twinning effect as discussed above, there actually exist two groups of excitation spectra symmetrized about $L=0$ that possess scattering intensities as shown in Fig.~\ref{fig3}(b).} To reproduce the magnetic excitation spectra observed in the twinned sample, we convolute the instrumental resolution with the calculated $S^{\alpha \beta}(\bm{Q},\omega)$ and superpose the $N$ and $R$ patterns of the excitations, which mainly affect the out-of-plane excitations. The final calculated magnon spectra of \MPS are presented in Figs.~\ref{fig3}(e)-\ref{fig3}(f). \new{For the  in-plane direction, the calculated excitation spectra successfully capture the shape of the measured spectra.} In addition, the dynamical spin structure factor along $\Gamma_1$ to $M_1$ is significantly smaller than that along $\Gamma_3$ to $M_2$ in the calculated spectra, which is in excellent agreement with the experimental observations shown in Fig.~\ref{fig3}(a). \new{Note that the calculated spectra exhibit weaker intensity at the top of the magnon band compared to the measured one. This discrepancy should stem from the mismatch between the resolution function and magnon dispersion in the INS experiment, which is not considered in the calculations. In terms of the out-of-plane spectra, our calculations  successfully capture the major features of the experimental spectra, especially the two winding bands around the band top. On the other hand, to reproduce the broadening of the measured spectra along the $L$ direction, one may need to take into account the two dimensionality.} These successful reproductions of the experimental data further validate the accuracy and reliability of our approach in characterizing the spin dynamics of \MPS.

\subsection{Magnon-phonon intersection}

In addition to magnons, our INS experiment also allows us to probe phonons. This capability enables us to examine whether there exist magnon polarons induced by the coupling between magnons and phonons in \MPS, as proposed in its sister compounds FePS$_3$ and FePSe$_3$ \cite{PhysRevLett.127.097401,MagnetoRamanStu,cui2023chirality}. A direct spectroscopic evidence of magnon polarons is the gap opening at the nominal intersection of the original magnon and phonon bands, accompanied by the mixing and interconversion of magnonic and phononic characters \cite{PhysRevLett.127.097401,MagnetoRamanStu,cui2023chirality, PhysRevLett.117.217205,PhysRevLett.123.167202,PhysRevLett.117.207203, NC14_6093, PhysRevLett.123.237207,PhysRevB.101.125111,PhysRevLett.124.147204}. Therefore, the key to addressing this issue lies in identifying the region where magnons and phonons intersect with each other. Figure~\ref{fig4}(a) shows the constant-energy ($E$) contour around 5~meV in the $(H,\,0,\,L)$ plane. Two rod-like excitations of magnons are clearly observed with negligible $L$ dependence, which reflects the quasi-2D nature of magnetic excitations due to the weak interlayer exchange interaction. Additionally, the oval-like excitations centered at $(0,\,0,\,6)$ is also visible, intersecting with the rod-like excitations in regions with $L$ around 4.7 and 7.3~rlu, as marked by the dashed lines in Fig.~\ref{fig4}(a). These two $L$s are symmetric about $L=6.0$~rlu. We attribute these distinct excitations with 3D nature to a phonon mode. Notably, the intensities of the rod-like excitations keep constant with increasing $L$, while those of the oval-like excitations increase [Fig.~\ref{fig4}(a)]. Such behaviors are consistent with the characteristics of the magnetic form factor corrected magnons and phonons, respectively.

To examine the intersection of the magnon and phonon dispersions, we select the region with $L=4.7$~rlu, where the scattering intensities of magnons and phonons are comparable [Fig.~\ref{fig4}(a)], and plot the excitation spectra along the [100] direction in Fig.~\ref{fig4}(b). In this plot, two distinct dispersions can be clearly identified within the same momentum-energy window: a V-shaped magnon dispersion starting from 3~meV and a W-shaped phonon dispersion distributed above 4.3~meV. The magnon and phonon dispersions appear to intersect at about $H=\pm 0.15$~rlu. However, within the instrumental resolution, there are no apparent gap openings at these intersections. To verify whether the magnon and phonon dispersions are distorted or renormalized due to the possible magnon-phonon coupling~\cite{PhysRevLett.117.217205,PhysRevLett.123.237207,PhysRevLett.123.167202,PhysRevB.101.125111,PhysRevLett.124.147204,PhysRevLett.117.207203,PhysRevLett.127.097401,MagnetoRamanStu,cui2023chirality,NC14_6093}, we investigate the individual magnon and phonon dispersions in Figs.~\ref{fig4}(c) and~\ref{fig4}(d), respectively. By deceasing $L$, we effectively suppress the scattering intensities of phonons, and we plot the pure magnons at a symmetrically equivalent position of $L=1.7=4.7-3.0$~rlu at 6~K in Fig.~\ref{fig4}(c). On the other hand, by increasing the temperature to 150~K, well above the $T_{\rm N}$, we obtain pure phonons at a higher intersection position with $L=7.3$~rlu in Fig.~\ref{fig4}(d). The momentum-energy relationships of the individual magnons [Fig.~\ref{fig4}(c)] and phonons [Fig.~\ref{fig4}(d)] are extracted by doing either constant-$\bm Q$ or constant-$E$ cuts through the excitation spectra. These extracted points are plotted over the spectra, with triangles and squares representing pure magnons and phonons, respectively, as shown in Fig.~\ref{fig4}(b). It is found that the extracted magnon and phonon bands match the overlapped magnon and phonon dispersions well, suggesting that there is no normalization between magnons and phonons. In other words, we find no obvious evidence of magnon polarons in \MPS.

\begin{figure}[htb]
	\centerline{\includegraphics[width=9.cm]{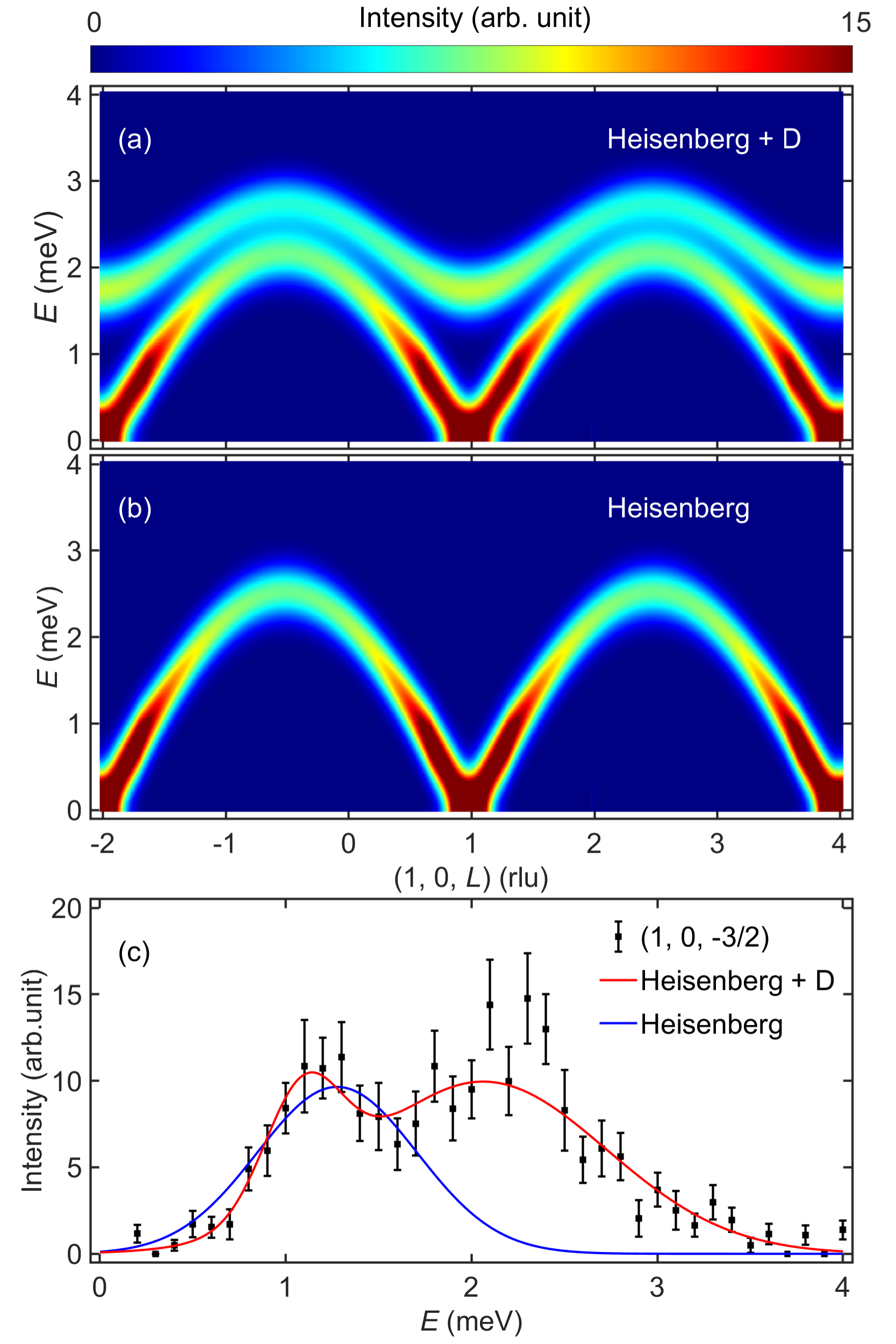}}
	\caption{{\new (a) and (b), Calculated magnon spectra with and without the $D$ term, respectively. (c), Comparison between the measured energy distribution of the spectral weight with the calculations with and without the $D$ term at ${\bm{Q}}=(1, 0, -3/2)$.}
	\label{fig5}}
\end{figure}

\section{Discussions}
\new{
\subsection{Easy-plane anisotropy}
We observe clear magnon dispersions in \MPS through INS experiments. To determine the effective spin model, we employ a Hamiltonian including Heisenberg term and easy-plane single-ion anisotropy ($D$) term to fit our INS data. Our theoretical fit gives a weak easy-plane anisotropy with the value of -0.035 meV. To estimate the significance of the easy-plane anisotropy and address the controversy regarding the effective spin model of \MPS~\cite{PJeevanandam_1999,ni2021imaging,PhysRevB.103.024414}, we fit the experimental data with Heisenberg term only. The exchange parameters given by the optimal fitting, are presented in the second row of Table~\ref{table1}. The goodness of the fitting $\chi ^{2}$ is \rnew{2.4}, larger than that with $D$ term. Table~\ref{table1} also lists the exchange parameters in a Heisenberg model from a previous INS measurement on powder samples~\cite{PhysRevB.103.024414}. It should be noted that since the definition of the spin Hamiltonian differs by a factor of 2, the parameters obtained from Ref.~\cite{PhysRevB.103.024414} are multiplied by 2 to keep consistency with our definition. We attempt to fit our data with their parameters, and the resulting value of $\chi ^{2}$ is \rnew{3.8}, also larger than the value with a $D$ term. Obviously, including the $D$ term yields a better fitting. More importantly, with and without the $D$ term result in qualitatively different magnon spectra. Figs.~\ref{fig5}(a) and \ref{fig5}(b) show the calculated magnon spectra with and without $D$ term, respectively. The twinning effect is not taken into consideration for brevity. The calculated spectra with $D$ term exhibit two magnon bands, including one acoustic and one optical branch as described previously, while the spectra without $D$ term only exhibit one magnon band dispersing from 0 meV to 3 meV, which makes it deviates from the measured spectra shown in Fig.~\ref{fig3}(b). To further illustrate this, we extract the measured energy distribution of the intensity at ${\bm Q}=(1, 0, -3/2)$ and plot it in Fig.~\ref{fig5}(c), which clearly shows two peaks at about 1.1 and 2.3~meV. Then, we fit the measured distribution with Heisenberg + $D$ model and Heisenberg model, respectively. It is apparent that the Heisenberg + $D$ model successfully captures the two-peak feature while the Heisenberg model fails to characterize the peak at 2.3~meV. Therefore, we believe that the easy-plane anisotropy, though weak, is necessary to be included to properly describe the spin dynamics of \MPS. The inclusion of this term is also consistent with the easy-plane magnetic order~\cite{PJeevanandam_1999,LEFLEM1982455,ni2021imaging}.}

\new{
\subsection{Connection with the XXZ model}
To take the easy-plane anisotropy into consideration, we can also adopt the following XXZ model: 
\begin{equation}\label{spinw}
	H=-\sum_{\langle i,j \rangle} J_{i,j}({S}^{x}_{i} {S}^{x}_{j}+{S}^{y}_{i} {S}^{y}_{j}+\delta_{i,j}{S}^{z}_{i}{S}^{z}_{j}).
\end{equation}
Here, the easy-plane anisotropic nature can be guaranteed by restricting the anisotropy factor $\delta_{i,j}$ to be smaller than 1. Nevertheless, within the framework of LSWT, a Heisenberg + $D$ model is equivalent to such an XXZ model in characterizing the spin dynamics~\cite{chen2024thermal,nolting2009quantum}. The impact of all anisotropy factors $\delta_{i,j}$ in the XXZ model can be equivalently displayed by the $D$ term in the Heisenberg + $D$ model. In the case of \MPS, these two models can be linked by 
\begin{equation}\label{spinw}
	D=\frac{3(1-\delta_{1})J_{1}+3(1-\delta_{2})J_{2}+3(1-\delta_{3})J_{3}+(1-\delta_{c})J_{c}}{2}.
\end{equation}
Here, $\delta_{i}$ with $\{i=1,2,3,c\}$ corresponds to the anisotropy factor of the bond that we concern in the Heisenberg~+~$D$ model. If we only consider a global anisotropy factor $\delta$, that is $\delta_{1}$=$\delta_{2}$=$\delta_{3}$=$\delta_{c}$=$\delta$, our refined $D$ term with the value of $-0.035$~meV gives rise to an anisotropy factor $\delta=0.981$~(Table~\ref{table1}). The corresponding parameters of the XXZ model are listed in fourth row of Table~\ref{table1}. At present, we believe the Heisenberg~+~$D$ or the equivalent XXZ model within the framework of LWST can already describe the magnetic excitation spectra well, but to further determine which one describes the spin dynamics of \MPS more precisely, one has to process the higher order term beyond LSWT~\cite{nolting2009quantum} or calculate the exchange parameters with first principles~\cite{scheie2023spin}.}

\new{
\subsection{Magnetic excitations in $\rm \textit{M}P\textit{X}_{3}$ compounds}
Now, we would like to briefly review the magnetic excitations of the $\rm \textit{M}P\textit{X}_{3}\ (M=Fe,\ Mn,\ Co,\ Ni; X=S,\ Se)$ compounds taking into account our results. The magnetic excitation spectra of FePS$_{3}$ and FePSe$_{3}$ exhibit similar magnon disperions with large bandwidth ($\gtrsim$20~meV) and excitation gap~($\sim15$~meV)~\cite{PhysRevB.94.214407,chen2024thermal}. This gap demonstrates the presence of a large easy-axis anisotropy and Ising nature of the spin model~\cite{PhysRevB.94.214407,chen2024thermal}. INS experiments on CoPS$_{3}$ observed 4 clear spin-wave branches with a bandwith of $\sim$18~meV and a large energy gap ($\gtrsim$14~meV), which can be well described by a model with Heisenberg term and biaxial single-ion anisotropy~\cite{PhysRevB.102.184429,PhysRevB.107.054438}. Moreover, it is found that the dominating exchange interactions is the third neighbor (3NN) antiferromagnetic interaction~\cite{PhysRevB.107.054438}. In the case of NiPS$_{3}$, 
INS experiments reported a wide magnetic excitation ($\gtrsim$45~meV) with a $\sim$1.3~meV excitation gap~\cite{scheie2023spin}. The LSWT fit also manifests a dominating 3NN exchange interaction, which stems from the large 3NN $e_{g}$-$e_{g}$ hopping integrals confirmed by $ab$-$initio$ density functional theory calculations~\cite{scheie2023spin}. As we can see, MnPSe$_{3}$ and MnPS$_{3}$ are the two compounds which have the narrowest magnon bandwidth ($\sim$8.5~meV and $\sim$11.3~meV~\cite{A_R_Wildes_1998}) and smallest excitaion gap (less than $\sim$0.1~meV~\cite{jana2023magnon} and $\sim$0.5~meV~\cite{A_R_Wildes_1998}) in the $\rm \textit{M}P\textit{X}_{3}$ material family. The weak spin-orbit coupling of Mn$^{2+}$ ions are potentially responsible for the weak exchange interaction and magnetic anisotropy. Note that Mn$^{2+}$ ions possess a half-filled 3$d$ shell, resulting in a quenched $L=0$ orbital moment.} 

\subsection{Absence of magnon polarons}
In addition to the clear magnon dispersions, we also observe the intersections between magnon and phonon dispersions. However, no obvious evidence of gap opening and hybridization between the two excitations has been identified in Fig.~\ref{fig4}. This suggests that there is negligible magnon-phonon coupling in \MPS, in contrast to the case in its sister compounds FePS$_3$ and FePSe$_3$~\cite{PhysRevLett.127.097401,MagnetoRamanStu,cui2023chirality}. Since the magnetic interactions depend on the relative ion positions, magnon-phonon coupling can arise due to the lattice vibrations. There are various mechanisms that magnon-phonon coupling can support magnon polarons, including magnetic dipolar interaction~\cite{PhysRevLett.117.217205}, magnetoelastic coupling~\cite{PhysRevLett.127.097401,PhysRevLett.123.237207, PhysRevLett.124.147204,PhysRevB.101.125111,RevModPhys.21.541, PhysRev.110.836}, and Dzyaloshinskii-Moriya (DM) interactions~\cite{PhysRevLett.123.167202, PhysRevB.105.L100402, ParKS}. While the strength of magnetic dipolar interaction is usually small compared to other interactions \cite{PhysRevB.93.014421}, its contribution is expected to be negligible in \MPS. On the other hand, magnetic anisotropy plays a crucial role in magnetoelastic coupling \cite{PhysRevLett.123.237207, PhysRevLett.124.147204,PhysRevLett.127.097401,PhysRevB.101.125111,RevModPhys.21.541, PhysRev.110.836}. However, in the case of \MPS, the magnetic anisotropy appears to be very weak, as inferred from the results of magnetic susceptibility [Fig.~\ref{fig1}(a)] and the excitation spectra [Fig.~\ref{fig3}(b)]. Therefore, the mechanism of magnetoelastic coupling is not applicable in \MPS. Additionally, the weak spin-orbit coupling results in the insignificance of the antisymmetric DM interaction in \MPS, making the DM mechanism insufficient to open a gap between magnon and phonon dispersions and generate the magnon polarons.

While our INS measurements do not directly detect magnon-phonon coupling or magnon polarons between one-magnon excitations and phonons, previous temperature-dependent magneto-Raman spectroscopy measurements have reported the presence of hybrid excitations involving two-magnon modes and optical phonon modes in \MPS~\cite{mai2021magnon}. This hybridization arises from the exchange striction effect, where the normal mode motion of the optical phonon mode modifies the distance between the nearest-neighbor and third-nearest neighbor Mn$^{2+}$ ions and affects the superexchange interactions $J_{1}$ and $J_{3}$ \cite{Baltensperger,mai2021magnon}. As the two-magnon modes and optical phonon modes overlap at the energy of about 16.2~meV, the coherent resonance of them induce the hybridization in \MPS. However, due to the weak intensity of the two-magnon excitation for a spin-$5/2$ system \cite{PhysRevB.72.014413}, our INS measurements do not capture the two-magnon modes, which should appear as a weak continuum-like excitation above the one-magnon dispersion in the INS spectra \cite{PhysRevB.72.014413,PhysRevLett.23.86,PhysRevB.94.104421,PhysRevLett.126.017201}. Consequently, the hybrid excitations of the two-magnon and phonon modes are not observed.

\section{Conclusions}
We conduct comprehensive INS measurements on single-crystal samples of \MPS to study the spin dynamics and determine the effective spin model. Clear magnon bands are observed in our INS experiment. \new{The excited magnon bands can be well described by an effective spin Hamiltonian with three intraplane exchange parameters ($J_{1}=-0.73$~meV, $J_{2}=-0.014$~meV, $J_{3}=-0.43$~meV), one interplane parameter ($J_{c}=-0.054$~meV), and an easy-plane single-ion anisotropy ($D=-0.035$~meV). This model is equivalent to an XXZ model with a global anisotropy factor $\delta=0.981$ in the framework of LSWT.} Furthermore, we observe that a quasi-2D magnon mode intersect with a phonon mode which disperse from $(0,0,6)$ rlu in the Brillouin zone. Nevertheless, none of the anomalous spectral features induced by magnon polarons have been identified in the intersection region, which we attribute to the weak spin-orbit coupling due to the quenched orbital momentum of~Mn$^{2+}$. 

\section{Acknowledgments}
	
We thank Jian-Xin Li, Shun-Li Yu, Zhao-Yang Dong, Wei Wang, Zhao-Long Gu and Li-Wei He for stimulating discussions. The work was supported by National Key Projects for Research and Development of China with Grant No.~2021YFA1400400, National Natural Science Foundation of China with Grant Nos.~12225407 and 12074174, China Postdoctoral Science Foundation with Grant Nos.~2022M711569 and 2022T150315, Jiangsu Province Excellent Postdoctoral Program with Grant No.~20220ZB5, and Fundamental Research Funds for the Central Universities. We acknowledge the neutron beam time from J-PARC with Proposal Nos.~2022A0040.


\begin{thebibliography}{70}%
	\makeatletter
	\providecommand \@ifxundefined [1]{%
		\@ifx{#1\undefined}
	}%
	\providecommand \@ifnum [1]{%
		\ifnum #1\expandafter \@firstoftwo
		\else \expandafter \@secondoftwo
		\fi
	}%
	\providecommand \@ifx [1]{%
		\ifx #1\expandafter \@firstoftwo
		\else \expandafter \@secondoftwo
		\fi
	}%
	\providecommand \natexlab [1]{#1}%
	\providecommand \enquote  [1]{``#1''}%
	\providecommand \bibnamefont  [1]{#1}%
	\providecommand \bibfnamefont [1]{#1}%
	\providecommand \citenamefont [1]{#1}%
	\providecommand \href@noop [0]{\@secondoftwo}%
	\providecommand \href [0]{\begingroup \@sanitize@url \@href}%
	\providecommand \@href[1]{\@@startlink{#1}\@@href}%
	\providecommand \@@href[1]{\endgroup#1\@@endlink}%
	\providecommand \@sanitize@url [0]{\catcode `\\12\catcode `\$12\catcode
		`\&12\catcode `\#12\catcode `\^12\catcode `\_12\catcode `\%12\relax}%
	\providecommand \@@startlink[1]{}%
	\providecommand \@@endlink[0]{}%
	\providecommand \url  [0]{\begingroup\@sanitize@url \@url }%
	\providecommand \@url [1]{\endgroup\@href {#1}{\urlprefix }}%
	\providecommand \urlprefix  [0]{URL }%
	\providecommand \Eprint [0]{\href }%
	\providecommand \doibase [0]{https://doi.org/}%
	\providecommand \selectlanguage [0]{\@gobble}%
	\providecommand \bibinfo  [0]{\@secondoftwo}%
	\providecommand \bibfield  [0]{\@secondoftwo}%
	\providecommand \translation [1]{[#1]}%
	\providecommand \BibitemOpen [0]{}%
	\providecommand \bibitemStop [0]{}%
	\providecommand \bibitemNoStop [0]{.\EOS\space}%
	\providecommand \EOS [0]{\spacefactor3000\relax}%
	\providecommand \BibitemShut  [1]{\csname bibitem#1\endcsname}%
	\let\auto@bib@innerbib\@empty
	\bibitem [{\citenamefont {Novoselov}\ \emph {et~al.}(2004)\citenamefont
		{Novoselov}, \citenamefont {Geim}, \citenamefont {Morozov}, \citenamefont
		{Jiang}, \citenamefont {Zhang}, \citenamefont {Dubonos}, \citenamefont
		{Grigorieva},\ and\ \citenamefont {Firsov}}]{Graphene}%
	\BibitemOpen
	\bibfield  {author} {\bibinfo {author} {\bibfnamefont {K.~S.}\ \bibnamefont
			{Novoselov}}, \bibinfo {author} {\bibfnamefont {A.~K.}\ \bibnamefont {Geim}},
		\bibinfo {author} {\bibfnamefont {S.~V.}\ \bibnamefont {Morozov}}, \bibinfo
		{author} {\bibfnamefont {D.}~\bibnamefont {Jiang}}, \bibinfo {author}
		{\bibfnamefont {Y.}~\bibnamefont {Zhang}}, \bibinfo {author} {\bibfnamefont
			{S.~V.}\ \bibnamefont {Dubonos}}, \bibinfo {author} {\bibfnamefont {I.~V.}\
			\bibnamefont {Grigorieva}},\ and\ \bibinfo {author} {\bibfnamefont {A.~A.}\
			\bibnamefont {Firsov}},\ }\bibfield  {title} {\bibinfo {title} {{Electric
				Field Effect in Atomically Thin Carbon Films}},\ }\href
	{https://doi.org/10.1126/science.1102896} {\bibfield  {journal} {\bibinfo
			{journal} {Science}\ }\textbf {\bibinfo {volume} {306}},\ \bibinfo {pages}
		{666} (\bibinfo {year} {2004})}\BibitemShut {NoStop}%
	\bibitem [{\citenamefont {D.~Sarma}\ \emph {et~al.}(2011)\citenamefont
		{D.~Sarma}, \citenamefont {Adam}, \citenamefont {Hwang},\ and\ \citenamefont
		{Rossi}}]{RevModPhys.83.407}%
	\BibitemOpen
	\bibfield  {author} {\bibinfo {author} {\bibfnamefont {S.}~\bibnamefont
			{D.~Sarma}}, \bibinfo {author} {\bibfnamefont {S.}~\bibnamefont {Adam}},
		\bibinfo {author} {\bibfnamefont {E.~H.}\ \bibnamefont {Hwang}},\ and\
		\bibinfo {author} {\bibfnamefont {E.}~\bibnamefont {Rossi}},\ }\bibfield
	{title} {\bibinfo {title} {{Electronic transport in two-dimensional
				graphene}},\ }\href {https://doi.org/10.1103/RevModPhys.83.407} {\bibfield
		{journal} {\bibinfo  {journal} {Rev. Mod. Phys.}\ }\textbf {\bibinfo {volume}
			{83}},\ \bibinfo {pages} {407} (\bibinfo {year} {2011})}\BibitemShut
	{NoStop}%
	\bibitem [{\citenamefont {Liu}\ \emph {et~al.}(2016)\citenamefont {Liu},
		\citenamefont {Weiss}, \citenamefont {Duan}, \citenamefont {Cheng},
		\citenamefont {Huang},\ and\ \citenamefont {Duan}}]{liu2016van}%
	\BibitemOpen
	\bibfield  {author} {\bibinfo {author} {\bibfnamefont {Y.}~\bibnamefont
			{Liu}}, \bibinfo {author} {\bibfnamefont {N.~O.}\ \bibnamefont {Weiss}},
		\bibinfo {author} {\bibfnamefont {X.}~\bibnamefont {Duan}}, \bibinfo {author}
		{\bibfnamefont {H.}~\bibnamefont {Cheng}}, \bibinfo {author} {\bibfnamefont
			{Y.}~\bibnamefont {Huang}},\ and\ \bibinfo {author} {\bibfnamefont
			{X.}~\bibnamefont {Duan}},\ }\bibfield  {title} {\bibinfo {title} {{Van der
				Waals heterostructures and devices}},\ }\href
	{https://doi.org/https://doi.org/10.1038/natrevmats.2016.42} {\bibfield
		{journal} {\bibinfo  {journal} {Nat. Rev. Mater.}\ }\textbf {\bibinfo
			{volume} {1}},\ \bibinfo {pages} {1} (\bibinfo {year} {2016})}\BibitemShut
	{NoStop}%
	\bibitem [{\citenamefont {Gu}\ \emph {et~al.}(2018)\citenamefont {Gu},
		\citenamefont {Wei}, \citenamefont {Yin}, \citenamefont {Li},\ and\
		\citenamefont {Yang}}]{RevModPhys.90.041002}%
	\BibitemOpen
	\bibfield  {author} {\bibinfo {author} {\bibfnamefont {X.}~\bibnamefont
			{Gu}}, \bibinfo {author} {\bibfnamefont {Y.}~\bibnamefont {Wei}}, \bibinfo
		{author} {\bibfnamefont {X.}~\bibnamefont {Yin}}, \bibinfo {author}
		{\bibfnamefont {B.}~\bibnamefont {Li}},\ and\ \bibinfo {author}
		{\bibfnamefont {R.}~\bibnamefont {Yang}},\ }\bibfield  {title} {\bibinfo
		{title} {{Colloquium: Phononic thermal properties of two-dimensional
				materials}},\ }\href {https://doi.org/10.1103/RevModPhys.90.041002}
	{\bibfield  {journal} {\bibinfo  {journal} {Rev. Mod. Phys.}\ }\textbf
		{\bibinfo {volume} {90}},\ \bibinfo {pages} {041002} (\bibinfo {year}
		{2018})}\BibitemShut {NoStop}%
	\bibitem [{\citenamefont {Avsar}\ \emph {et~al.}(2020)\citenamefont {Avsar},
		\citenamefont {Ochoa}, \citenamefont {Guinea}, \citenamefont {\"Ozyilmaz},
		\citenamefont {van Wees},\ and\ \citenamefont
		{Vera-Marun}}]{RevModPhys.92.021003}%
	\BibitemOpen
	\bibfield  {author} {\bibinfo {author} {\bibfnamefont {A.}~\bibnamefont
			{Avsar}}, \bibinfo {author} {\bibfnamefont {H.}~\bibnamefont {Ochoa}},
		\bibinfo {author} {\bibfnamefont {F.}~\bibnamefont {Guinea}}, \bibinfo
		{author} {\bibfnamefont {B.}~\bibnamefont {\"Ozyilmaz}}, \bibinfo {author}
		{\bibfnamefont {B.~J.}\ \bibnamefont {van Wees}},\ and\ \bibinfo {author}
		{\bibfnamefont {I.~J.}\ \bibnamefont {Vera-Marun}},\ }\bibfield  {title}
	{\bibinfo {title} {{Colloquium: Spintronics in graphene and other
				two-dimensional materials}},\ }\href
	{https://doi.org/10.1103/RevModPhys.92.021003} {\bibfield  {journal}
		{\bibinfo  {journal} {Rev. Mod. Phys.}\ }\textbf {\bibinfo {volume} {92}},\
		\bibinfo {pages} {021003} (\bibinfo {year} {2020})}\BibitemShut {NoStop}%
	\bibitem [{\citenamefont {Huang}\ \emph {et~al.}(2017)\citenamefont {Huang},
		\citenamefont {Clark}, \citenamefont {Navarro-Moratalla}, \citenamefont
		{Klein}, \citenamefont {Cheng}, \citenamefont {Seyler}, \citenamefont
		{Zhong}, \citenamefont {Schmidgall}, \citenamefont {McGuire}, \citenamefont
		{Cobden} \emph {et~al.}}]{huang2017layer}%
	\BibitemOpen
	\bibfield  {author} {\bibinfo {author} {\bibfnamefont {B.}~\bibnamefont
			{Huang}}, \bibinfo {author} {\bibfnamefont {G.}~\bibnamefont {Clark}},
		\bibinfo {author} {\bibfnamefont {E.}~\bibnamefont {Navarro-Moratalla}},
		\bibinfo {author} {\bibfnamefont {D.~R.}\ \bibnamefont {Klein}}, \bibinfo
		{author} {\bibfnamefont {R.}~\bibnamefont {Cheng}}, \bibinfo {author}
		{\bibfnamefont {K.~L.}\ \bibnamefont {Seyler}}, \bibinfo {author}
		{\bibfnamefont {D.}~\bibnamefont {Zhong}}, \bibinfo {author} {\bibfnamefont
			{E.}~\bibnamefont {Schmidgall}}, \bibinfo {author} {\bibfnamefont {M.~A.}\
			\bibnamefont {McGuire}}, \bibinfo {author} {\bibfnamefont {D.~H.}\
			\bibnamefont {Cobden}}, \emph {et~al.},\ }\bibfield  {title} {\bibinfo
		{title} {{Layer-dependent ferromagnetism in a van der Waals crystal down to
				the monolayer limit}},\ }\href
	{https://doi.org/https://doi.org/10.1038/nature22391} {\bibfield  {journal}
		{\bibinfo  {journal} {Nature}\ }\textbf {\bibinfo {volume} {546}},\ \bibinfo
		{pages} {270} (\bibinfo {year} {2017})}\BibitemShut {NoStop}%
	\bibitem [{\citenamefont {Gong}\ \emph {et~al.}(2017)\citenamefont {Gong},
		\citenamefont {Li}, \citenamefont {Li}, \citenamefont {Ji}, \citenamefont
		{Stern}, \citenamefont {Xia}, \citenamefont {Cao}, \citenamefont {Bao},
		\citenamefont {Wang}, \citenamefont {Wang} \emph
		{et~al.}}]{gong2017discovery}%
	\BibitemOpen
	\bibfield  {author} {\bibinfo {author} {\bibfnamefont {C.}~\bibnamefont
			{Gong}}, \bibinfo {author} {\bibfnamefont {L.}~\bibnamefont {Li}}, \bibinfo
		{author} {\bibfnamefont {Z.}~\bibnamefont {Li}}, \bibinfo {author}
		{\bibfnamefont {H.}~\bibnamefont {Ji}}, \bibinfo {author} {\bibfnamefont
			{A.}~\bibnamefont {Stern}}, \bibinfo {author} {\bibfnamefont
			{Y.}~\bibnamefont {Xia}}, \bibinfo {author} {\bibfnamefont {T.}~\bibnamefont
			{Cao}}, \bibinfo {author} {\bibfnamefont {W.}~\bibnamefont {Bao}}, \bibinfo
		{author} {\bibfnamefont {C.}~\bibnamefont {Wang}}, \bibinfo {author}
		{\bibfnamefont {Y.}~\bibnamefont {Wang}}, \emph {et~al.},\ }\bibfield
	{title} {\bibinfo {title} {{Discovery of intrinsic ferromagnetism in
				two-dimensional van der Waals crystals}},\ }\href
	{https://doi.org/https://doi.org/10.1038/nature22060} {\bibfield  {journal}
		{\bibinfo  {journal} {Nature}\ }\textbf {\bibinfo {volume} {546}},\ \bibinfo
		{pages} {265} (\bibinfo {year} {2017})}\BibitemShut {NoStop}%
	\bibitem [{\citenamefont {Bedoya-Pinto}\ \emph {et~al.}(2021)\citenamefont
		{Bedoya-Pinto}, \citenamefont {Ji}, \citenamefont {Pandeya}, \citenamefont
		{Gargiani}, \citenamefont {Valvidares}, \citenamefont {Sessi}, \citenamefont
		{Taylor}, \citenamefont {Radu}, \citenamefont {Chang},\ and\ \citenamefont
		{Parkin}}]{Intrinsic2DXY}%
	\BibitemOpen
	\bibfield  {author} {\bibinfo {author} {\bibfnamefont {A.}~\bibnamefont
			{Bedoya-Pinto}}, \bibinfo {author} {\bibfnamefont {J.}~\bibnamefont {Ji}},
		\bibinfo {author} {\bibfnamefont {A.~K.}\ \bibnamefont {Pandeya}}, \bibinfo
		{author} {\bibfnamefont {P.}~\bibnamefont {Gargiani}}, \bibinfo {author}
		{\bibfnamefont {M.}~\bibnamefont {Valvidares}}, \bibinfo {author}
		{\bibfnamefont {P.}~\bibnamefont {Sessi}}, \bibinfo {author} {\bibfnamefont
			{J.~M.}\ \bibnamefont {Taylor}}, \bibinfo {author} {\bibfnamefont
			{F.}~\bibnamefont {Radu}}, \bibinfo {author} {\bibfnamefont {K.}~\bibnamefont
			{Chang}},\ and\ \bibinfo {author} {\bibfnamefont {S.~S.~P.}\ \bibnamefont
			{Parkin}},\ }\bibfield  {title} {\bibinfo {title} {{Intrinsic 2D-XY
				ferromagnetism in a van der Waals monolayer}},\ }\href
	{https://doi.org/10.1126/science.abd5146} {\bibfield  {journal} {\bibinfo
			{journal} {Science}\ }\textbf {\bibinfo {volume} {374}},\ \bibinfo {pages}
		{616} (\bibinfo {year} {2021})}\BibitemShut {NoStop}%
	\bibitem [{\citenamefont {Mermin}\ and\ \citenamefont
		{Wagner}(1966)}]{PhysRevLett.17.1133}%
	\BibitemOpen
	\bibfield  {author} {\bibinfo {author} {\bibfnamefont {N.~D.}\ \bibnamefont
			{Mermin}}\ and\ \bibinfo {author} {\bibfnamefont {H.}~\bibnamefont
			{Wagner}},\ }\bibfield  {title} {\bibinfo {title} {{Absence of Ferromagnetism
				or Antiferromagnetism in One- or Two-Dimensional Isotropic Heisenberg
				Models}},\ }\href {https://doi.org/10.1103/PhysRevLett.17.1133} {\bibfield
		{journal} {\bibinfo  {journal} {Phys. Rev. Lett.}\ }\textbf {\bibinfo
			{volume} {17}},\ \bibinfo {pages} {1133} (\bibinfo {year}
		{1966})}\BibitemShut {NoStop}%
	\bibitem [{\citenamefont {Deng}\ \emph {et~al.}(2018)\citenamefont {Deng},
		\citenamefont {Yu}, \citenamefont {Song}, \citenamefont {Zhang},
		\citenamefont {Wang}, \citenamefont {Sun}, \citenamefont {Yi}, \citenamefont
		{Wu}, \citenamefont {Wu}, \citenamefont {Zhu} \emph {et~al.}}]{deng2018gate}%
	\BibitemOpen
	\bibfield  {author} {\bibinfo {author} {\bibfnamefont {Y.}~\bibnamefont
			{Deng}}, \bibinfo {author} {\bibfnamefont {Y.}~\bibnamefont {Yu}}, \bibinfo
		{author} {\bibfnamefont {Y.}~\bibnamefont {Song}}, \bibinfo {author}
		{\bibfnamefont {J.}~\bibnamefont {Zhang}}, \bibinfo {author} {\bibfnamefont
			{N.~Z.}\ \bibnamefont {Wang}}, \bibinfo {author} {\bibfnamefont
			{Z.}~\bibnamefont {Sun}}, \bibinfo {author} {\bibfnamefont {Y.}~\bibnamefont
			{Yi}}, \bibinfo {author} {\bibfnamefont {Y.~Z.}\ \bibnamefont {Wu}}, \bibinfo
		{author} {\bibfnamefont {S.}~\bibnamefont {Wu}}, \bibinfo {author}
		{\bibfnamefont {J.}~\bibnamefont {Zhu}}, \emph {et~al.},\ }\bibfield  {title}
	{\bibinfo {title} {{Gate-tunable room-temperature ferromagnetism in
				two-dimensional Fe$_3$GeTe$_2$}},\ }\href
	{https://doi.org/doi.org/10.1038/s41586-018-0626-9} {\bibfield  {journal}
		{\bibinfo  {journal} {Nature}\ }\textbf {\bibinfo {volume} {563}},\ \bibinfo
		{pages} {94} (\bibinfo {year} {2018})}\BibitemShut {NoStop}%
	\bibitem [{\citenamefont {Bao}\ \emph {et~al.}(2022)\citenamefont {Bao},
		\citenamefont {Wang}, \citenamefont {Shangguan}, \citenamefont {Cai},
		\citenamefont {Dong}, \citenamefont {Huang}, \citenamefont {Si},
		\citenamefont {Ma}, \citenamefont {Kajimoto}, \citenamefont {Ikeuchi},
		\citenamefont {Yano}, \citenamefont {Yu}, \citenamefont {Wan}, \citenamefont
		{Li},\ and\ \citenamefont {Wen}}]{PhysRevX.12.011022}%
	\BibitemOpen
	\bibfield  {author} {\bibinfo {author} {\bibfnamefont {S.}~\bibnamefont
			{Bao}}, \bibinfo {author} {\bibfnamefont {W.}~\bibnamefont {Wang}}, \bibinfo
		{author} {\bibfnamefont {Y.}~\bibnamefont {Shangguan}}, \bibinfo {author}
		{\bibfnamefont {Z.}~\bibnamefont {Cai}}, \bibinfo {author} {\bibfnamefont
			{Z.-Y.}\ \bibnamefont {Dong}}, \bibinfo {author} {\bibfnamefont
			{Z.}~\bibnamefont {Huang}}, \bibinfo {author} {\bibfnamefont
			{W.}~\bibnamefont {Si}}, \bibinfo {author} {\bibfnamefont {Z.}~\bibnamefont
			{Ma}}, \bibinfo {author} {\bibfnamefont {R.}~\bibnamefont {Kajimoto}},
		\bibinfo {author} {\bibfnamefont {K.}~\bibnamefont {Ikeuchi}}, \bibinfo
		{author} {\bibfnamefont {S.-i.}\ \bibnamefont {Yano}}, \bibinfo {author}
		{\bibfnamefont {S.-L.}\ \bibnamefont {Yu}}, \bibinfo {author} {\bibfnamefont
			{X.}~\bibnamefont {Wan}}, \bibinfo {author} {\bibfnamefont {J.-X.}\
			\bibnamefont {Li}},\ and\ \bibinfo {author} {\bibfnamefont {J.}~\bibnamefont
			{Wen}},\ }\bibfield  {title} {\bibinfo {title} {{Neutron Spectroscopy
				Evidence on the Dual Nature of Magnetic Excitations in a van der Waals
				Metallic Ferromagnet ${\mathrm{Fe}}_{2.72}{\mathrm{GeTe}}_{2}$}},\ }\href
	{https://doi.org/10.1103/PhysRevX.12.011022} {\bibfield  {journal} {\bibinfo
			{journal} {Phys. Rev. X}\ }\textbf {\bibinfo {volume} {12}},\ \bibinfo
		{pages} {011022} (\bibinfo {year} {2022})}\BibitemShut {NoStop}%
	\bibitem [{\citenamefont {Lee}\ \emph {et~al.}(2016)\citenamefont {Lee},
		\citenamefont {Lee}, \citenamefont {Ryoo}, \citenamefont {Kang},
		\citenamefont {Kim}, \citenamefont {Kim}, \citenamefont {Park}, \citenamefont
		{Park},\ and\ \citenamefont {Cheong}}]{lee2016ising}%
	\BibitemOpen
	\bibfield  {author} {\bibinfo {author} {\bibfnamefont {J.}~\bibnamefont
			{Lee}}, \bibinfo {author} {\bibfnamefont {S.}~\bibnamefont {Lee}}, \bibinfo
		{author} {\bibfnamefont {J.~H.}\ \bibnamefont {Ryoo}}, \bibinfo {author}
		{\bibfnamefont {S.}~\bibnamefont {Kang}}, \bibinfo {author} {\bibfnamefont
			{T.~Y.}\ \bibnamefont {Kim}}, \bibinfo {author} {\bibfnamefont
			{P.}~\bibnamefont {Kim}}, \bibinfo {author} {\bibfnamefont {C.}~\bibnamefont
			{Park}}, \bibinfo {author} {\bibfnamefont {J.}~\bibnamefont {Park}},\ and\
		\bibinfo {author} {\bibfnamefont {H.}~\bibnamefont {Cheong}},\ }\bibfield
	{title} {\bibinfo {title} {{Ising-Type Magnetic Ordering in Atomically Thin
				${\mathrm{FePS}}_{3}$}},\ }\href
	{https://doi.org/10.1021/acs.nanolett.6b03052} {\bibfield  {journal}
		{\bibinfo  {journal} {Nano Lett.}\ }\textbf {\bibinfo {volume} {16}},\
		\bibinfo {pages} {7433} (\bibinfo {year} {2016})}\BibitemShut {NoStop}%
	\bibitem [{\citenamefont {Ni}\ \emph {et~al.}(2021)\citenamefont {Ni},
		\citenamefont {Haglund}, \citenamefont {Wang}, \citenamefont {Xu},
		\citenamefont {Bernhard}, \citenamefont {Mandrus}, \citenamefont {Qian},
		\citenamefont {Mele}, \citenamefont {Kane},\ and\ \citenamefont
		{Wu}}]{ni2021imaging}%
	\BibitemOpen
	\bibfield  {author} {\bibinfo {author} {\bibfnamefont {Z.}~\bibnamefont
			{Ni}}, \bibinfo {author} {\bibfnamefont {A.}~\bibnamefont {Haglund}},
		\bibinfo {author} {\bibfnamefont {H.}~\bibnamefont {Wang}}, \bibinfo {author}
		{\bibfnamefont {B.}~\bibnamefont {Xu}}, \bibinfo {author} {\bibfnamefont
			{C.}~\bibnamefont {Bernhard}}, \bibinfo {author} {\bibfnamefont
			{D.}~\bibnamefont {Mandrus}}, \bibinfo {author} {\bibfnamefont
			{X.}~\bibnamefont {Qian}}, \bibinfo {author} {\bibfnamefont {E.}~\bibnamefont
			{Mele}}, \bibinfo {author} {\bibfnamefont {C.}~\bibnamefont {Kane}},\ and\
		\bibinfo {author} {\bibfnamefont {L.}~\bibnamefont {Wu}},\ }\bibfield
	{title} {\bibinfo {title} {{Imaging the N{\'e}el vector switching in the
				monolayer antiferromagnet ${\mathrm{MnPSe}}_{3}$ with strain-controlled Ising
				order}},\ }\href {https://doi.org/https://doi.org/10.1038/s41565-021-00885-5}
	{\bibfield  {journal} {\bibinfo  {journal} {Nat. Nanotechnol.}\ }\textbf
		{\bibinfo {volume} {16}},\ \bibinfo {pages} {782} (\bibinfo {year}
		{2021})}\BibitemShut {NoStop}%
	\bibitem [{\citenamefont {Klingen}\ \emph {et~al.}(1968)\citenamefont
		{Klingen}, \citenamefont {Eulenberger},\ and\ \citenamefont
		{Hahn}}]{klingen1968hexathio}%
	\BibitemOpen
	\bibfield  {author} {\bibinfo {author} {\bibfnamefont {W.}~\bibnamefont
			{Klingen}}, \bibinfo {author} {\bibfnamefont {G.}~\bibnamefont
			{Eulenberger}},\ and\ \bibinfo {author} {\bibfnamefont {H.}~\bibnamefont
			{Hahn}},\ }\bibfield  {title} {\bibinfo {title} {{About Hexathio- and
				Hexaselenohypodiphosphate Type ${\mathrm{M_{2}^{II}P_{2}X_{6}}}$}},\ }\href
	{https://doi.org/https://doi.org/10.1007/BF00606219} {\bibfield  {journal}
		{\bibinfo  {journal} {Nat. Sci.}\ }\textbf {\bibinfo {volume} {55}},\
		\bibinfo {pages} {229} (\bibinfo {year} {1968})}\BibitemShut {NoStop}%
	\bibitem [{\citenamefont {Taylor}\ \emph {et~al.}(1974)\citenamefont {Taylor},
		\citenamefont {Steger}, \citenamefont {Wold},\ and\ \citenamefont
		{Kostiner}}]{Prep}%
	\BibitemOpen
	\bibfield  {author} {\bibinfo {author} {\bibfnamefont {B.}~\bibnamefont
			{Taylor}}, \bibinfo {author} {\bibfnamefont {J.}~\bibnamefont {Steger}},
		\bibinfo {author} {\bibfnamefont {A.}~\bibnamefont {Wold}},\ and\ \bibinfo
		{author} {\bibfnamefont {E.}~\bibnamefont {Kostiner}},\ }\bibfield  {title}
	{\bibinfo {title} {{Preparation and properties of iron phosphorus
				triselenide, ${\mathrm{FePSe}}_{3}$}},\ }\href
	{https://doi.org/10.1021/ic50141a034} {\bibfield  {journal} {\bibinfo
			{journal} {Inorg. Chem.}\ }\textbf {\bibinfo {volume} {13}},\ \bibinfo
		{pages} {2719} (\bibinfo {year} {1974})}\BibitemShut {NoStop}%
	\bibitem [{\citenamefont {Wiedenmann}\ \emph {et~al.}(1981)\citenamefont
		{Wiedenmann}, \citenamefont {Rossat-Mignod}, \citenamefont {Louisy},
		\citenamefont {Brec},\ and\ \citenamefont {Rouxel}}]{WIEDENMANN19811067}%
	\BibitemOpen
	\bibfield  {author} {\bibinfo {author} {\bibfnamefont {A.}~\bibnamefont
			{Wiedenmann}}, \bibinfo {author} {\bibfnamefont {J.}~\bibnamefont
			{Rossat-Mignod}}, \bibinfo {author} {\bibfnamefont {A.}~\bibnamefont
			{Louisy}}, \bibinfo {author} {\bibfnamefont {R.}~\bibnamefont {Brec}},\ and\
		\bibinfo {author} {\bibfnamefont {J.}~\bibnamefont {Rouxel}},\ }\bibfield
	{title} {\bibinfo {title} {{Neutron diffraction study of the layered
				compounds ${\mathrm{MnPSe}}_{3}$ and ${\mathrm{FePSe}}_{3}$}},\ }\href
	{https://doi.org/https://doi.org/10.1016/0038-1098(81)90253-2} {\bibfield
		{journal} {\bibinfo  {journal} {Solid State Commun.}\ }\textbf {\bibinfo
			{volume} {40}},\ \bibinfo {pages} {1067} (\bibinfo {year}
		{1981})}\BibitemShut {NoStop}%
	\bibitem [{\citenamefont {Kurosawa}\ \emph {et~al.}(1983)\citenamefont
		{Kurosawa}, \citenamefont {Saito},\ and\ \citenamefont {Yamaguchi}}]{NeutDi}%
	\BibitemOpen
	\bibfield  {author} {\bibinfo {author} {\bibfnamefont {K.}~\bibnamefont
			{Kurosawa}}, \bibinfo {author} {\bibfnamefont {S.}~\bibnamefont {Saito}},\
		and\ \bibinfo {author} {\bibfnamefont {Y.}~\bibnamefont {Yamaguchi}},\
	}\bibfield  {title} {\bibinfo {title} {{Neutron Diffraction Study on
				${\mathrm{MnPS}}_{3}$ and ${\mathrm{FePS}}_{3}$}},\ }\href
	{https://doi.org/10.1143/JPSJ.52.3919} {\bibfield  {journal} {\bibinfo
			{journal} {J. Phys. Soc. Jpn}\ }\textbf {\bibinfo {volume} {52}},\ \bibinfo
		{pages} {3919} (\bibinfo {year} {1983})}\BibitemShut {NoStop}%
	\bibitem [{\citenamefont {Okuda}\ \emph {et~al.}(1986)\citenamefont {Okuda},
		\citenamefont {Kurosawa}, \citenamefont {Saito}, \citenamefont {Honda},
		\citenamefont {Yu},\ and\ \citenamefont {Date}}]{MagneticPro}%
	\BibitemOpen
	\bibfield  {author} {\bibinfo {author} {\bibfnamefont {K.}~\bibnamefont
			{Okuda}}, \bibinfo {author} {\bibfnamefont {K.}~\bibnamefont {Kurosawa}},
		\bibinfo {author} {\bibfnamefont {S.}~\bibnamefont {Saito}}, \bibinfo
		{author} {\bibfnamefont {M.}~\bibnamefont {Honda}}, \bibinfo {author}
		{\bibfnamefont {Z.}~\bibnamefont {Yu}},\ and\ \bibinfo {author}
		{\bibfnamefont {M.}~\bibnamefont {Date}},\ }\bibfield  {title} {\bibinfo
		{title} {{Magnetic Properties of Layered Compound ${\mathrm{MnPS}}_{3}$}},\
	}\href {https://doi.org/10.1143/JPSJ.55.4456} {\bibfield  {journal} {\bibinfo
			{journal} {J. Phys. Soc. Jpn}\ }\textbf {\bibinfo {volume} {55}},\ \bibinfo
		{pages} {4456} (\bibinfo {year} {1986})}\BibitemShut {NoStop}%
	\bibitem [{\citenamefont {Joy}\ and\ \citenamefont
		{Vasudevan}(1992)}]{PhysRevB.46.5425}%
	\BibitemOpen
	\bibfield  {author} {\bibinfo {author} {\bibfnamefont {P.~A.}\ \bibnamefont
			{Joy}}\ and\ \bibinfo {author} {\bibfnamefont {S.}~\bibnamefont
			{Vasudevan}},\ }\bibfield  {title} {\bibinfo {title} {{Magnetism in the
				layered transition-metal thiophosphates ${\mathrm{MPS}}_{3}$ (M=Mn, Fe, and
				Ni)}},\ }\href {https://doi.org/10.1103/PhysRevB.46.5425} {\bibfield
		{journal} {\bibinfo  {journal} {Phys. Rev. B}\ }\textbf {\bibinfo {volume}
			{46}},\ \bibinfo {pages} {5425} (\bibinfo {year} {1992})}\BibitemShut
	{NoStop}%
	\bibitem [{\citenamefont {Wildes}\ \emph {et~al.}(1994)\citenamefont {Wildes},
		\citenamefont {Kennedy},\ and\ \citenamefont {Hicks}}]{A_R_Wildes_1994}%
	\BibitemOpen
	\bibfield  {author} {\bibinfo {author} {\bibfnamefont {A.~R.}\ \bibnamefont
			{Wildes}}, \bibinfo {author} {\bibfnamefont {S.~J.}\ \bibnamefont
			{Kennedy}},\ and\ \bibinfo {author} {\bibfnamefont {T.~J.}\ \bibnamefont
			{Hicks}},\ }\bibfield  {title} {\bibinfo {title} {{True two-dimensional
				magnetic ordering in ${\mathrm{MnPS}}_{3}$}},\ }\href
	{https://doi.org/10.1088/0953-8984/6/24/002} {\bibfield  {journal} {\bibinfo
			{journal} {J. Phys. Condens. Matter}\ }\textbf {\bibinfo {volume} {6}},\
		\bibinfo {pages} {L335} (\bibinfo {year} {1994})}\BibitemShut {NoStop}%
	\bibitem [{\citenamefont {Wildes}\ \emph {et~al.}(1998)\citenamefont {Wildes},
		\citenamefont {Roessli}, \citenamefont {Lebech},\ and\ \citenamefont
		{Godfrey}}]{A_R_Wildes_1998}%
	\BibitemOpen
	\bibfield  {author} {\bibinfo {author} {\bibfnamefont {A.~R.}\ \bibnamefont
			{Wildes}}, \bibinfo {author} {\bibfnamefont {B.}~\bibnamefont {Roessli}},
		\bibinfo {author} {\bibfnamefont {B.}~\bibnamefont {Lebech}},\ and\ \bibinfo
		{author} {\bibfnamefont {K.~W.}\ \bibnamefont {Godfrey}},\ }\bibfield
	{title} {\bibinfo {title} {{Spin waves and the critical behaviour of the
				magnetization in ${\mathrm{MnPS}}_{3}$}},\ }\href
	{https://doi.org/10.1088/0953-8984/10/28/020} {\bibfield  {journal} {\bibinfo
			{journal} {J. Phys. Condens. Matter}\ }\textbf {\bibinfo {volume} {10}},\
		\bibinfo {pages} {6417} (\bibinfo {year} {1998})}\BibitemShut {NoStop}%
	\bibitem [{\citenamefont {Wildes}\ \emph {et~al.}(2006)\citenamefont {Wildes},
		\citenamefont {R\o{}nnow}, \citenamefont {Roessli}, \citenamefont {Harris},\
		and\ \citenamefont {Godfrey}}]{PhysRevB.74.094422}%
	\BibitemOpen
	\bibfield  {author} {\bibinfo {author} {\bibfnamefont {A.~R.}\ \bibnamefont
			{Wildes}}, \bibinfo {author} {\bibfnamefont {H.~M.}\ \bibnamefont
			{R\o{}nnow}}, \bibinfo {author} {\bibfnamefont {B.}~\bibnamefont {Roessli}},
		\bibinfo {author} {\bibfnamefont {M.~J.}\ \bibnamefont {Harris}},\ and\
		\bibinfo {author} {\bibfnamefont {K.~W.}\ \bibnamefont {Godfrey}},\
	}\bibfield  {title} {\bibinfo {title} {{Static and dynamic critical
				properties of the quasi-two-dimensional antiferromagnet
				${\mathrm{MnPS}}_{3}$}},\ }\href {https://doi.org/10.1103/PhysRevB.74.094422}
	{\bibfield  {journal} {\bibinfo  {journal} {Phys. Rev. B}\ }\textbf {\bibinfo
			{volume} {74}},\ \bibinfo {pages} {094422} (\bibinfo {year}
		{2006})}\BibitemShut {NoStop}%
	\bibitem [{\citenamefont {Rule}\ \emph {et~al.}(2007)\citenamefont {Rule},
		\citenamefont {McIntyre}, \citenamefont {Kennedy},\ and\ \citenamefont
		{Hicks}}]{PhysRevB.76.134402}%
	\BibitemOpen
	\bibfield  {author} {\bibinfo {author} {\bibfnamefont {K.~C.}\ \bibnamefont
			{Rule}}, \bibinfo {author} {\bibfnamefont {G.~J.}\ \bibnamefont {McIntyre}},
		\bibinfo {author} {\bibfnamefont {S.~J.}\ \bibnamefont {Kennedy}},\ and\
		\bibinfo {author} {\bibfnamefont {T.~J.}\ \bibnamefont {Hicks}},\ }\bibfield
	{title} {\bibinfo {title} {{Single-crystal and powder neutron diffraction
				experiments on $\mathrm{Fe}\mathrm{P}{\mathrm{S}}_{3}$: Search for the
				magnetic structure}},\ }\href {https://doi.org/10.1103/PhysRevB.76.134402}
	{\bibfield  {journal} {\bibinfo  {journal} {Phys. Rev. B}\ }\textbf {\bibinfo
			{volume} {76}},\ \bibinfo {pages} {134402} (\bibinfo {year}
		{2007})}\BibitemShut {NoStop}%
	\bibitem [{\citenamefont {Wildes}\ \emph {et~al.}(2012)\citenamefont {Wildes},
		\citenamefont {Rule}, \citenamefont {Bewley}, \citenamefont {Enderle},\ and\
		\citenamefont {Hicks}}]{Wildes_2012}%
	\BibitemOpen
	\bibfield  {author} {\bibinfo {author} {\bibfnamefont {A.~R.}\ \bibnamefont
			{Wildes}}, \bibinfo {author} {\bibfnamefont {K.~C.}\ \bibnamefont {Rule}},
		\bibinfo {author} {\bibfnamefont {R.~I.}\ \bibnamefont {Bewley}}, \bibinfo
		{author} {\bibfnamefont {M.}~\bibnamefont {Enderle}},\ and\ \bibinfo {author}
		{\bibfnamefont {T.~J.}\ \bibnamefont {Hicks}},\ }\bibfield  {title} {\bibinfo
		{title} {{The magnon dynamics and spin exchange parameters of
				${\mathrm{FePS}}_{3}$}},\ }\href
	{https://doi.org/10.1088/0953-8984/24/41/416004} {\bibfield  {journal}
		{\bibinfo  {journal} {J. Phys. Condens. Matter}\ }\textbf {\bibinfo {volume}
			{24}},\ \bibinfo {pages} {416004} (\bibinfo {year} {2012})}\BibitemShut
	{NoStop}%
	\bibitem [{\citenamefont {Lan\ifmmode~\mbox{\c{c}}\else \c{c}\fi{}on}\ \emph
		{et~al.}(2016)\citenamefont {Lan\ifmmode~\mbox{\c{c}}\else \c{c}\fi{}on},
		\citenamefont {Walker}, \citenamefont {Ressouche}, \citenamefont {Ouladdiaf},
		\citenamefont {Rule}, \citenamefont {McIntyre}, \citenamefont {Hicks},
		\citenamefont {R\o{}nnow},\ and\ \citenamefont
		{Wildes}}]{PhysRevB.94.214407}%
	\BibitemOpen
	\bibfield  {author} {\bibinfo {author} {\bibfnamefont {D.}~\bibnamefont
			{Lan\ifmmode~\mbox{\c{c}}\else \c{c}\fi{}on}}, \bibinfo {author}
		{\bibfnamefont {H.~C.}\ \bibnamefont {Walker}}, \bibinfo {author}
		{\bibfnamefont {E.}~\bibnamefont {Ressouche}}, \bibinfo {author}
		{\bibfnamefont {B.}~\bibnamefont {Ouladdiaf}}, \bibinfo {author}
		{\bibfnamefont {K.~C.}\ \bibnamefont {Rule}}, \bibinfo {author}
		{\bibfnamefont {G.~J.}\ \bibnamefont {McIntyre}}, \bibinfo {author}
		{\bibfnamefont {T.~J.}\ \bibnamefont {Hicks}}, \bibinfo {author}
		{\bibfnamefont {H.~M.}\ \bibnamefont {R\o{}nnow}},\ and\ \bibinfo {author}
		{\bibfnamefont {A.~R.}\ \bibnamefont {Wildes}},\ }\bibfield  {title}
	{\bibinfo {title} {{Magnetic structure and magnon dynamics of the
				quasi-two-dimensional antiferromagnet ${\mathrm{FePS}}_{3}$}},\ }\href
	{https://doi.org/10.1103/PhysRevB.94.214407} {\bibfield  {journal} {\bibinfo
			{journal} {Phys. Rev. B}\ }\textbf {\bibinfo {volume} {94}},\ \bibinfo
		{pages} {214407} (\bibinfo {year} {2016})}\BibitemShut {NoStop}%
	\bibitem [{\citenamefont {Chen}\ \emph {et~al.}(2024)\citenamefont {Chen},
		\citenamefont {Teng}, \citenamefont {Hu}, \citenamefont {Ye}, \citenamefont
		{Granroth}, \citenamefont {Yi}, \citenamefont {Chung}, \citenamefont
		{Birgeneau},\ and\ \citenamefont {Dai}}]{chen2024thermal}%
	\BibitemOpen
	\bibfield  {author} {\bibinfo {author} {\bibfnamefont {L.}~\bibnamefont
			{Chen}}, \bibinfo {author} {\bibfnamefont {X.}~\bibnamefont {Teng}}, \bibinfo
		{author} {\bibfnamefont {D.}~\bibnamefont {Hu}}, \bibinfo {author}
		{\bibfnamefont {F.}~\bibnamefont {Ye}}, \bibinfo {author} {\bibfnamefont
			{G.~E.}\ \bibnamefont {Granroth}}, \bibinfo {author} {\bibfnamefont
			{M.}~\bibnamefont {Yi}}, \bibinfo {author} {\bibfnamefont {J.-H.}\
			\bibnamefont {Chung}}, \bibinfo {author} {\bibfnamefont {R.~J.}\ \bibnamefont
			{Birgeneau}},\ and\ \bibinfo {author} {\bibfnamefont {P.}~\bibnamefont
			{Dai}},\ }\href@noop {} {\bibinfo {title} {{Thermal evolution of spin
				excitations in honeycomb Ising antiferromagnetic FePSe$_{3}$}}} \href
	{https://doi.org/10.1038/s41535-024-00651-5} {\bibfield  {journal} {\bibinfo
			{journal} {npj Quantum Mater.}\ }\textbf {\bibinfo {volume} {9}},\ \bibinfo
		{pages} {40} (\bibinfo {year} {2024})}\BibitemShut {NoStop}%
	\bibitem [{\citenamefont {Wildes}\ \emph {et~al.}(2023)\citenamefont {Wildes},
		\citenamefont {F\aa{}k}, \citenamefont {Hansen}, \citenamefont {Enderle},
		\citenamefont {Stewart}, \citenamefont {Testa}, \citenamefont {R\o{}nnow},
		\citenamefont {Kim},\ and\ \citenamefont {Park}}]{PhysRevB.107.054438}%
	\BibitemOpen
	\bibfield  {author} {\bibinfo {author} {\bibfnamefont {A.~R.}\ \bibnamefont
			{Wildes}}, \bibinfo {author} {\bibfnamefont {B.}~\bibnamefont {F\aa{}k}},
		\bibinfo {author} {\bibfnamefont {U.~B.}\ \bibnamefont {Hansen}}, \bibinfo
		{author} {\bibfnamefont {M.}~\bibnamefont {Enderle}}, \bibinfo {author}
		{\bibfnamefont {J.~R.}\ \bibnamefont {Stewart}}, \bibinfo {author}
		{\bibfnamefont {L.}~\bibnamefont {Testa}}, \bibinfo {author} {\bibfnamefont
			{H.~M.}\ \bibnamefont {R\o{}nnow}}, \bibinfo {author} {\bibfnamefont
			{C.}~\bibnamefont {Kim}},\ and\ \bibinfo {author} {\bibfnamefont {J.-G.}\
			\bibnamefont {Park}},\ }\bibfield  {title} {\bibinfo {title} {{Spin wave
				spectra of single crystal ${\mathrm{CoPS}}_{3}$}},\ }\href
	{https://doi.org/10.1103/PhysRevB.107.054438} {\bibfield  {journal} {\bibinfo
			{journal} {Phys. Rev. B}\ }\textbf {\bibinfo {volume} {107}},\ \bibinfo
		{pages} {054438} (\bibinfo {year} {2023})}\BibitemShut {NoStop}%
	\bibitem [{\citenamefont {Kim}\ \emph {et~al.}(2020)\citenamefont {Kim},
		\citenamefont {Jeong}, \citenamefont {Park}, \citenamefont {Masuda},
		\citenamefont {Asai}, \citenamefont {Itoh}, \citenamefont {Kim},
		\citenamefont {Wildes},\ and\ \citenamefont {Park}}]{PhysRevB.102.184429}%
	\BibitemOpen
	\bibfield  {author} {\bibinfo {author} {\bibfnamefont {C.}~\bibnamefont
			{Kim}}, \bibinfo {author} {\bibfnamefont {J.}~\bibnamefont {Jeong}}, \bibinfo
		{author} {\bibfnamefont {P.}~\bibnamefont {Park}}, \bibinfo {author}
		{\bibfnamefont {T.}~\bibnamefont {Masuda}}, \bibinfo {author} {\bibfnamefont
			{S.}~\bibnamefont {Asai}}, \bibinfo {author} {\bibfnamefont {S.}~\bibnamefont
			{Itoh}}, \bibinfo {author} {\bibfnamefont {H.-S.}\ \bibnamefont {Kim}},
		\bibinfo {author} {\bibfnamefont {A.}~\bibnamefont {Wildes}},\ and\ \bibinfo
		{author} {\bibfnamefont {J.}~\bibnamefont {Park}},\ }\bibfield  {title}
	{\bibinfo {title} {{Spin waves in the two-dimensional honeycomb lattice
				XXZ-type van der Waals antiferromagnet ${\mathrm{CoPS}}_{3}$}},\ }\href
	{https://doi.org/10.1103/PhysRevB.102.184429} {\bibfield  {journal} {\bibinfo
			{journal} {Phys. Rev. B}\ }\textbf {\bibinfo {volume} {102}},\ \bibinfo
		{pages} {184429} (\bibinfo {year} {2020})}\BibitemShut {NoStop}%
	\bibitem [{\citenamefont {Scheie}\ \emph {et~al.}(2023)\citenamefont {Scheie},
		\citenamefont {Park}, \citenamefont {Villanova}, \citenamefont {Granroth},
		\citenamefont {Sarkis}, \citenamefont {Zhang}, \citenamefont {Stone},
		\citenamefont {Park}, \citenamefont {Okamoto}, \citenamefont {Berlijn},\ and\
		\citenamefont {Tennant}}]{scheie2023spin}%
	\BibitemOpen
	\bibfield  {author} {\bibinfo {author} {\bibfnamefont {A.}~\bibnamefont
			{Scheie}}, \bibinfo {author} {\bibfnamefont {P.}~\bibnamefont {Park}},
		\bibinfo {author} {\bibfnamefont {J.~W.}\ \bibnamefont {Villanova}}, \bibinfo
		{author} {\bibfnamefont {G.~E.}\ \bibnamefont {Granroth}}, \bibinfo {author}
		{\bibfnamefont {C.~L.}\ \bibnamefont {Sarkis}}, \bibinfo {author}
		{\bibfnamefont {H.}~\bibnamefont {Zhang}}, \bibinfo {author} {\bibfnamefont
			{M.~B.}\ \bibnamefont {Stone}}, \bibinfo {author} {\bibfnamefont {J.-G.}\
			\bibnamefont {Park}}, \bibinfo {author} {\bibfnamefont {S.}~\bibnamefont
			{Okamoto}}, \bibinfo {author} {\bibfnamefont {T.}~\bibnamefont {Berlijn}},\
		and\ \bibinfo {author} {\bibfnamefont {D.~A.}\ \bibnamefont {Tennant}},\
	}\bibfield  {title} {\bibinfo {title} {{Spin wave Hamiltonian and anomalous
				scattering in ${\mathrm{NiPS}}_{3}$}},\ }\href
	{https://doi.org/10.1103/PhysRevB.108.104402} {\bibfield  {journal} {\bibinfo
			{journal} {Phys. Rev. B}\ }\textbf {\bibinfo {volume} {108}},\ \bibinfo
		{pages} {104402} (\bibinfo {year} {2023})}\BibitemShut {NoStop}%
	\bibitem [{\citenamefont {Wildes}\ \emph {et~al.}(2015)\citenamefont {Wildes},
		\citenamefont {Simonet}, \citenamefont {Ressouche}, \citenamefont {McIntyre},
		\citenamefont {Avdeev}, \citenamefont {Suard}, \citenamefont {Kimber},
		\citenamefont {Lan\ifmmode~\mbox{\c{c}}\else \c{c}\fi{}on}, \citenamefont
		{Pepe}, \citenamefont {Moubaraki},\ and\ \citenamefont
		{Hicks}}]{PhysRevB.92.224408}%
	\BibitemOpen
	\bibfield  {author} {\bibinfo {author} {\bibfnamefont {A.~R.}\ \bibnamefont
			{Wildes}}, \bibinfo {author} {\bibfnamefont {V.}~\bibnamefont {Simonet}},
		\bibinfo {author} {\bibfnamefont {E.}~\bibnamefont {Ressouche}}, \bibinfo
		{author} {\bibfnamefont {G.~J.}\ \bibnamefont {McIntyre}}, \bibinfo {author}
		{\bibfnamefont {M.}~\bibnamefont {Avdeev}}, \bibinfo {author} {\bibfnamefont
			{E.}~\bibnamefont {Suard}}, \bibinfo {author} {\bibfnamefont {S.~A.~J.}\
			\bibnamefont {Kimber}}, \bibinfo {author} {\bibfnamefont {D.}~\bibnamefont
			{Lan\ifmmode~\mbox{\c{c}}\else \c{c}\fi{}on}}, \bibinfo {author}
		{\bibfnamefont {G.}~\bibnamefont {Pepe}}, \bibinfo {author} {\bibfnamefont
			{B.}~\bibnamefont {Moubaraki}},\ and\ \bibinfo {author} {\bibfnamefont
			{T.~J.}\ \bibnamefont {Hicks}},\ }\bibfield  {title} {\bibinfo {title}
		{{Magnetic structure of the quasi-two-dimensional antiferromagnet
				${\text{NiPS}}_{3}$}},\ }\href {https://doi.org/10.1103/PhysRevB.92.224408}
	{\bibfield  {journal} {\bibinfo  {journal} {Phys. Rev. B}\ }\textbf {\bibinfo
			{volume} {92}},\ \bibinfo {pages} {224408} (\bibinfo {year}
		{2015})}\BibitemShut {NoStop}%
	\bibitem [{\citenamefont {Bhutani}\ \emph {et~al.}(2020)\citenamefont
		{Bhutani}, \citenamefont {Zuo}, \citenamefont {McAuliffe}, \citenamefont
		{dela Cruz},\ and\ \citenamefont {Shoemaker}}]{PhysRevMaterials.4.034411}%
	\BibitemOpen
	\bibfield  {author} {\bibinfo {author} {\bibfnamefont {A.}~\bibnamefont
			{Bhutani}}, \bibinfo {author} {\bibfnamefont {J.~L.}\ \bibnamefont {Zuo}},
		\bibinfo {author} {\bibfnamefont {R.~D.}\ \bibnamefont {McAuliffe}}, \bibinfo
		{author} {\bibfnamefont {C.~R.}\ \bibnamefont {dela Cruz}},\ and\ \bibinfo
		{author} {\bibfnamefont {D.~P.}\ \bibnamefont {Shoemaker}},\ }\bibfield
	{title} {\bibinfo {title} {{Strong anisotropy in the mixed antiferromagnetic
				system
				${\mathrm{Mn}}_{1\ensuremath{-}x}{\mathrm{Fe}}_{x}{\mathrm{PSe}}_{3}$}},\
	}\href {https://doi.org/10.1103/PhysRevMaterials.4.034411} {\bibfield
		{journal} {\bibinfo  {journal} {Phys. Rev. Mater.}\ }\textbf {\bibinfo
			{volume} {4}},\ \bibinfo {pages} {034411} (\bibinfo {year}
		{2020})}\BibitemShut {NoStop}%
	\bibitem [{\citenamefont {Selter}\ \emph {et~al.}(2021)\citenamefont {Selter},
		\citenamefont {Shemerliuk}, \citenamefont {Sturza}, \citenamefont {Wolter},
		\citenamefont {B\"uchner},\ and\ \citenamefont
		{Aswartham}}]{PhysRevMaterials.5.073401}%
	\BibitemOpen
	\bibfield  {author} {\bibinfo {author} {\bibfnamefont {S.}~\bibnamefont
			{Selter}}, \bibinfo {author} {\bibfnamefont {Y.}~\bibnamefont {Shemerliuk}},
		\bibinfo {author} {\bibfnamefont {M.-I.}\ \bibnamefont {Sturza}}, \bibinfo
		{author} {\bibfnamefont {A.~U.~B.}\ \bibnamefont {Wolter}}, \bibinfo {author}
		{\bibfnamefont {B.}~\bibnamefont {B\"uchner}},\ and\ \bibinfo {author}
		{\bibfnamefont {S.}~\bibnamefont {Aswartham}},\ }\bibfield  {title} {\bibinfo
		{title} {{Crystal growth and anisotropic magnetic properties of
				quasi-two-dimensional
				${(\mathrm{Fe}}_{1\ensuremath{-}x}{\mathrm{Ni}}_{x}{)}_{2}{\mathrm{P}}_{2}{\mathrm{S}}_{6}$}},\
	}\href {https://doi.org/10.1103/PhysRevMaterials.5.073401} {\bibfield
		{journal} {\bibinfo  {journal} {Phys. Rev. Mater.}\ }\textbf {\bibinfo
			{volume} {5}},\ \bibinfo {pages} {073401} (\bibinfo {year}
		{2021})}\BibitemShut {NoStop}%
	\bibitem [{\citenamefont {Coak}\ \emph {et~al.}(2021)\citenamefont {Coak},
		\citenamefont {Jarvis}, \citenamefont {Hamidov}, \citenamefont {Wildes},
		\citenamefont {Paddison}, \citenamefont {Liu}, \citenamefont {Haines},
		\citenamefont {Dang}, \citenamefont {Kichanov}, \citenamefont {Savenko},
		\citenamefont {Lee}, \citenamefont {Kratochv\'{\i}lov\'a}, \citenamefont
		{Klotz}, \citenamefont {Hansen}, \citenamefont {Kozlenko}, \citenamefont
		{Park},\ and\ \citenamefont {Saxena}}]{PhysRevX.11.011024}%
	\BibitemOpen
	\bibfield  {author} {\bibinfo {author} {\bibfnamefont {M.~J.}\ \bibnamefont
			{Coak}}, \bibinfo {author} {\bibfnamefont {D.~M.}\ \bibnamefont {Jarvis}},
		\bibinfo {author} {\bibfnamefont {H.}~\bibnamefont {Hamidov}}, \bibinfo
		{author} {\bibfnamefont {A.~R.}\ \bibnamefont {Wildes}}, \bibinfo {author}
		{\bibfnamefont {J.~A.~M.}\ \bibnamefont {Paddison}}, \bibinfo {author}
		{\bibfnamefont {C.}~\bibnamefont {Liu}}, \bibinfo {author} {\bibfnamefont
			{C.~R.~S.}\ \bibnamefont {Haines}}, \bibinfo {author} {\bibfnamefont {N.~T.}\
			\bibnamefont {Dang}}, \bibinfo {author} {\bibfnamefont {S.~E.}\ \bibnamefont
			{Kichanov}}, \bibinfo {author} {\bibfnamefont {B.~N.}\ \bibnamefont
			{Savenko}}, \bibinfo {author} {\bibfnamefont {S.}~\bibnamefont {Lee}},
		\bibinfo {author} {\bibfnamefont {M.}~\bibnamefont {Kratochv\'{\i}lov\'a}},
		\bibinfo {author} {\bibfnamefont {S.}~\bibnamefont {Klotz}}, \bibinfo
		{author} {\bibfnamefont {T.~C.}\ \bibnamefont {Hansen}}, \bibinfo {author}
		{\bibfnamefont {D.~P.}\ \bibnamefont {Kozlenko}}, \bibinfo {author}
		{\bibfnamefont {J.}~\bibnamefont {Park}},\ and\ \bibinfo {author}
		{\bibfnamefont {S.~S.}\ \bibnamefont {Saxena}},\ }\bibfield  {title}
	{\bibinfo {title} {{Emergent Magnetic Phases in Pressure-Tuned van der Waals
				Antiferromagnet ${\mathrm{FePS}}_{3}$}},\ }\href
	{https://doi.org/10.1103/PhysRevX.11.011024} {\bibfield  {journal} {\bibinfo
			{journal} {Phys. Rev. X}\ }\textbf {\bibinfo {volume} {11}},\ \bibinfo
		{pages} {011024} (\bibinfo {year} {2021})}\BibitemShut {NoStop}%
	\bibitem [{\citenamefont {Ma}\ \emph {et~al.}(2021)\citenamefont {Ma},
		\citenamefont {Wang}, \citenamefont {Yin}, \citenamefont {Yue}, \citenamefont
		{Dai}, \citenamefont {Cheng}, \citenamefont {Ji}, \citenamefont {Jin},
		\citenamefont {Hong}, \citenamefont {Wang} \emph
		{et~al.}}]{ma2021dimensional}%
	\BibitemOpen
	\bibfield  {author} {\bibinfo {author} {\bibfnamefont {X.}~\bibnamefont
			{Ma}}, \bibinfo {author} {\bibfnamefont {Y.}~\bibnamefont {Wang}}, \bibinfo
		{author} {\bibfnamefont {Y.}~\bibnamefont {Yin}}, \bibinfo {author}
		{\bibfnamefont {B.}~\bibnamefont {Yue}}, \bibinfo {author} {\bibfnamefont
			{J.}~\bibnamefont {Dai}}, \bibinfo {author} {\bibfnamefont {J.}~\bibnamefont
			{Cheng}}, \bibinfo {author} {\bibfnamefont {J.}~\bibnamefont {Ji}}, \bibinfo
		{author} {\bibfnamefont {F.}~\bibnamefont {Jin}}, \bibinfo {author}
		{\bibfnamefont {F.}~\bibnamefont {Hong}}, \bibinfo {author} {\bibfnamefont
			{J.}~\bibnamefont {Wang}}, \emph {et~al.},\ }\bibfield  {title} {\bibinfo
		{title} {{Dimensional crossover tuned by pressure in layered magnetic
				${\mathrm{NiPS}}_{3}$}},\ }\href
	{https://doi.org/https://doi.org/10.1007/s11433-021-1727-6} {\bibfield
		{journal} {\bibinfo  {journal} {Sci. China Phys., Mech.}\ }\textbf {\bibinfo
			{volume} {64}},\ \bibinfo {pages} {297011} (\bibinfo {year}
		{2021})}\BibitemShut {NoStop}%
	\bibitem [{\citenamefont {Kang}\ \emph {et~al.}(2020)\citenamefont {Kang},
		\citenamefont {Kim}, \citenamefont {Kim}, \citenamefont {Kim}, \citenamefont
		{Sim}, \citenamefont {Lee}, \citenamefont {Lee}, \citenamefont {Park},
		\citenamefont {Yun}, \citenamefont {Kim} \emph {et~al.}}]{kang2020coherent}%
	\BibitemOpen
	\bibfield  {author} {\bibinfo {author} {\bibfnamefont {S.}~\bibnamefont
			{Kang}}, \bibinfo {author} {\bibfnamefont {K.}~\bibnamefont {Kim}}, \bibinfo
		{author} {\bibfnamefont {B.~H.}\ \bibnamefont {Kim}}, \bibinfo {author}
		{\bibfnamefont {J.}~\bibnamefont {Kim}}, \bibinfo {author} {\bibfnamefont
			{K.~I.}\ \bibnamefont {Sim}}, \bibinfo {author} {\bibfnamefont
			{J.}~\bibnamefont {Lee}}, \bibinfo {author} {\bibfnamefont {S.}~\bibnamefont
			{Lee}}, \bibinfo {author} {\bibfnamefont {K.}~\bibnamefont {Park}}, \bibinfo
		{author} {\bibfnamefont {S.}~\bibnamefont {Yun}}, \bibinfo {author}
		{\bibfnamefont {T.}~\bibnamefont {Kim}}, \emph {et~al.},\ }\bibfield  {title}
	{\bibinfo {title} {{Coherent many-body exciton in van der Waals
				antiferromagnet ${\mathrm{NiPS}}_{3}$}},\ }\href
	{https://doi.org/https://doi.org/10.1038/s41586-020-2520-5} {\bibfield
		{journal} {\bibinfo  {journal} {Nature}\ }\textbf {\bibinfo {volume} {583}},\
		\bibinfo {pages} {785} (\bibinfo {year} {2020})}\BibitemShut {NoStop}%
	\bibitem [{\citenamefont {Hwangbo}\ \emph {et~al.}(2021)\citenamefont
		{Hwangbo}, \citenamefont {Zhang}, \citenamefont {Jiang}, \citenamefont
		{Wang}, \citenamefont {Fonseca}, \citenamefont {Wang}, \citenamefont
		{Diederich}, \citenamefont {Gamelin}, \citenamefont {Xiao}, \citenamefont
		{Chu} \emph {et~al.}}]{hwangbo2021highly}%
	\BibitemOpen
	\bibfield  {author} {\bibinfo {author} {\bibfnamefont {K.}~\bibnamefont
			{Hwangbo}}, \bibinfo {author} {\bibfnamefont {Q.}~\bibnamefont {Zhang}},
		\bibinfo {author} {\bibfnamefont {Q.}~\bibnamefont {Jiang}}, \bibinfo
		{author} {\bibfnamefont {Y.}~\bibnamefont {Wang}}, \bibinfo {author}
		{\bibfnamefont {J.}~\bibnamefont {Fonseca}}, \bibinfo {author} {\bibfnamefont
			{C.}~\bibnamefont {Wang}}, \bibinfo {author} {\bibfnamefont {G.~M.}\
			\bibnamefont {Diederich}}, \bibinfo {author} {\bibfnamefont {D.~R.}\
			\bibnamefont {Gamelin}}, \bibinfo {author} {\bibfnamefont {D.}~\bibnamefont
			{Xiao}}, \bibinfo {author} {\bibfnamefont {J.}~\bibnamefont {Chu}}, \emph
		{et~al.},\ }\bibfield  {title} {\bibinfo {title} {{Highly anisotropic
				excitons and multiple phonon bound states in a van der Waals
				antiferromagnetic insulator}},\ }\href
	{https://doi.org/https://doi.org/10.1038/s41565-021-00873-9} {\bibfield
		{journal} {\bibinfo  {journal} {Nat. Nanotechnol.}\ }\textbf {\bibinfo
			{volume} {16}},\ \bibinfo {pages} {655} (\bibinfo {year} {2021})}\BibitemShut
	{NoStop}%
	\bibitem [{\citenamefont {Liu}\ \emph {et~al.}(2021)\citenamefont {Liu},
		\citenamefont {Granados~del \'Aguila}, \citenamefont {Bhowmick},
		\citenamefont {Gan}, \citenamefont {Thu Ha~Do}, \citenamefont {Prosnikov},
		\citenamefont {Sedmidubsk\'y}, \citenamefont {Sofer}, \citenamefont
		{Christianen}, \citenamefont {Sengupta},\ and\ \citenamefont
		{Xiong}}]{PhysRevLett.127.097401}%
	\BibitemOpen
	\bibfield  {author} {\bibinfo {author} {\bibfnamefont {S.}~\bibnamefont
			{Liu}}, \bibinfo {author} {\bibfnamefont {A.}~\bibnamefont {Granados~del
				\'Aguila}}, \bibinfo {author} {\bibfnamefont {D.}~\bibnamefont {Bhowmick}},
		\bibinfo {author} {\bibfnamefont {C.~K.}\ \bibnamefont {Gan}}, \bibinfo
		{author} {\bibfnamefont {T.}~\bibnamefont {Thu Ha~Do}}, \bibinfo {author}
		{\bibfnamefont {M.~A.}\ \bibnamefont {Prosnikov}}, \bibinfo {author}
		{\bibfnamefont {D.}~\bibnamefont {Sedmidubsk\'y}}, \bibinfo {author}
		{\bibfnamefont {Z.}~\bibnamefont {Sofer}}, \bibinfo {author} {\bibfnamefont
			{P.~C.~M.}\ \bibnamefont {Christianen}}, \bibinfo {author} {\bibfnamefont
			{P.}~\bibnamefont {Sengupta}},\ and\ \bibinfo {author} {\bibfnamefont
			{Q.}~\bibnamefont {Xiong}},\ }\bibfield  {title} {\bibinfo {title} {{Direct
				Observation of Magnon-Phonon Strong Coupling in Two-Dimensional
				Antiferromagnet at High Magnetic Fields}},\ }\href
	{https://doi.org/10.1103/PhysRevLett.127.097401} {\bibfield  {journal}
		{\bibinfo  {journal} {Phys. Rev. Lett.}\ }\textbf {\bibinfo {volume} {127}},\
		\bibinfo {pages} {097401} (\bibinfo {year} {2021})}\BibitemShut {NoStop}%
	\bibitem [{\citenamefont {Sun}\ \emph {et~al.}(2022)\citenamefont {Sun},
		\citenamefont {Lai}, \citenamefont {Pang}, \citenamefont {Liu}, \citenamefont
		{Tan},\ and\ \citenamefont {Zhang}}]{MagnetoRamanStu}%
	\BibitemOpen
	\bibfield  {author} {\bibinfo {author} {\bibfnamefont {Y.}~\bibnamefont
			{Sun}}, \bibinfo {author} {\bibfnamefont {J.}~\bibnamefont {Lai}}, \bibinfo
		{author} {\bibfnamefont {S.}~\bibnamefont {Pang}}, \bibinfo {author}
		{\bibfnamefont {X.}~\bibnamefont {Liu}}, \bibinfo {author} {\bibfnamefont
			{P.}~\bibnamefont {Tan}},\ and\ \bibinfo {author} {\bibfnamefont
			{J.}~\bibnamefont {Zhang}},\ }\bibfield  {title} {\bibinfo {title}
		{{Magneto-Raman Study of Magnon-Phonon Coupling in Two-Dimensional Ising
				Antiferromagnetic ${\mathrm{FePS}}_{3}$}},\ }\href
	{https://doi.org/10.1021/acs.jpclett.2c00023} {\bibfield  {journal} {\bibinfo
			{journal} {J. Phys. Chem. Lett.}\ }\textbf {\bibinfo {volume} {13}},\
		\bibinfo {pages} {1533} (\bibinfo {year} {2022})}\BibitemShut {NoStop}%
	\bibitem [{\citenamefont {Cui}\ \emph {et~al.}(2023)\citenamefont {Cui},
		\citenamefont {Bostr{\"o}m}, \citenamefont {Ozerov}, \citenamefont {Wu},
		\citenamefont {Jiang}, \citenamefont {Chu}, \citenamefont {Li}, \citenamefont
		{Liu}, \citenamefont {Xu}, \citenamefont {Rubio} \emph
		{et~al.}}]{cui2023chirality}%
	\BibitemOpen
	\bibfield  {author} {\bibinfo {author} {\bibfnamefont {J.}~\bibnamefont
			{Cui}}, \bibinfo {author} {\bibfnamefont {E.~V.}\ \bibnamefont
			{Bostr{\"o}m}}, \bibinfo {author} {\bibfnamefont {M.}~\bibnamefont {Ozerov}},
		\bibinfo {author} {\bibfnamefont {F.}~\bibnamefont {Wu}}, \bibinfo {author}
		{\bibfnamefont {Q.}~\bibnamefont {Jiang}}, \bibinfo {author} {\bibfnamefont
			{J.}~\bibnamefont {Chu}}, \bibinfo {author} {\bibfnamefont {C.}~\bibnamefont
			{Li}}, \bibinfo {author} {\bibfnamefont {F.}~\bibnamefont {Liu}}, \bibinfo
		{author} {\bibfnamefont {X.}~\bibnamefont {Xu}}, \bibinfo {author}
		{\bibfnamefont {A.}~\bibnamefont {Rubio}}, \emph {et~al.},\ }\bibfield
	{title} {\bibinfo {title} {{Chirality selective magnon-phonon hybridization
				and magnon-induced chiral phonons in a layered zigzag antiferromagnet}},\
	}\href {https://doi.org/https://doi.org/10.1038/s41467-023-39123-y}
	{\bibfield  {journal} {\bibinfo  {journal} {Nat. Commun.}\ }\textbf {\bibinfo
			{volume} {14}},\ \bibinfo {pages} {3396} (\bibinfo {year}
		{2023})}\BibitemShut {NoStop}%
	\bibitem [{\citenamefont {{Le Flem}}\ \emph {et~al.}(1982)\citenamefont {{Le
				Flem}}, \citenamefont {Brec}, \citenamefont {Ouvard}, \citenamefont
		{Louisy},\ and\ \citenamefont {Segransan}}]{LEFLEM1982455}%
	\BibitemOpen
	\bibfield  {author} {\bibinfo {author} {\bibfnamefont {G.}~\bibnamefont {{Le
					Flem}}}, \bibinfo {author} {\bibfnamefont {R.}~\bibnamefont {Brec}}, \bibinfo
		{author} {\bibfnamefont {G.}~\bibnamefont {Ouvard}}, \bibinfo {author}
		{\bibfnamefont {A.}~\bibnamefont {Louisy}},\ and\ \bibinfo {author}
		{\bibfnamefont {P.}~\bibnamefont {Segransan}},\ }\bibfield  {title} {\bibinfo
		{title} {{Magnetic interactions in the layer compounds ${\mathrm{MPX}}_{3}$
				(M = Mn, Fe, Ni; X = S, Se)}},\ }\href
	{https://doi.org/https://doi.org/10.1016/0022-3697(82)90156-1} {\bibfield
		{journal} {\bibinfo  {journal} {J. Phys. Chem. Solids}\ }\textbf {\bibinfo
			{volume} {43}},\ \bibinfo {pages} {455} (\bibinfo {year} {1982})}\BibitemShut
	{NoStop}%
	\bibitem [{\citenamefont {Jeevanandam}\ and\ \citenamefont
		{Vasudevan}(1999)}]{PJeevanandam_1999}%
	\BibitemOpen
	\bibfield  {author} {\bibinfo {author} {\bibfnamefont {P.}~\bibnamefont
			{Jeevanandam}}\ and\ \bibinfo {author} {\bibfnamefont {S.}~\bibnamefont
			{Vasudevan}},\ }\bibfield  {title} {\bibinfo {title} {{Magnetism in
				${\mathrm{MnPSe}}_{3}$: a layered 3d5 antiferromagnet with unusually large XY
				anisotropy}},\ }\href {https://doi.org/10.1088/0953-8984/11/17/314}
	{\bibfield  {journal} {\bibinfo  {journal} {J. Phys. Condens. Matter}\
		}\textbf {\bibinfo {volume} {11}},\ \bibinfo {pages} {3563} (\bibinfo {year}
		{1999})}\BibitemShut {NoStop}%
	\bibitem [{\citenamefont {Calder}\ \emph {et~al.}(2021)\citenamefont {Calder},
		\citenamefont {Haglund}, \citenamefont {Kolesnikov},\ and\ \citenamefont
		{Mandrus}}]{PhysRevB.103.024414}%
	\BibitemOpen
	\bibfield  {author} {\bibinfo {author} {\bibfnamefont {S.}~\bibnamefont
			{Calder}}, \bibinfo {author} {\bibfnamefont {A.~V.}\ \bibnamefont {Haglund}},
		\bibinfo {author} {\bibfnamefont {A.~I.}\ \bibnamefont {Kolesnikov}},\ and\
		\bibinfo {author} {\bibfnamefont {D.}~\bibnamefont {Mandrus}},\ }\bibfield
	{title} {\bibinfo {title} {{Magnetic exchange interactions in the van der
				Waals layered antiferromagnet $\mathrm{Mn}\mathrm{P}{\mathrm{Se}}_{3}$}},\
	}\href {https://doi.org/10.1103/PhysRevB.103.024414} {\bibfield  {journal}
		{\bibinfo  {journal} {Phys. Rev. B}\ }\textbf {\bibinfo {volume} {103}},\
		\bibinfo {pages} {024414} (\bibinfo {year} {2021})}\BibitemShut {NoStop}%
	\bibitem [{\citenamefont {Mai}\ \emph {et~al.}(2021)\citenamefont {Mai},
		\citenamefont {Garrity}, \citenamefont {McCreary}, \citenamefont {Argo},
		\citenamefont {Simpson}, \citenamefont {Doan-Nguyen}, \citenamefont
		{Aguilar},\ and\ \citenamefont {Walker}}]{mai2021magnon}%
	\BibitemOpen
	\bibfield  {author} {\bibinfo {author} {\bibfnamefont {T.~T.}\ \bibnamefont
			{Mai}}, \bibinfo {author} {\bibfnamefont {K.~F.}\ \bibnamefont {Garrity}},
		\bibinfo {author} {\bibfnamefont {A.}~\bibnamefont {McCreary}}, \bibinfo
		{author} {\bibfnamefont {J.}~\bibnamefont {Argo}}, \bibinfo {author}
		{\bibfnamefont {J.~R.}\ \bibnamefont {Simpson}}, \bibinfo {author}
		{\bibfnamefont {V.}~\bibnamefont {Doan-Nguyen}}, \bibinfo {author}
		{\bibfnamefont {R.~V.}\ \bibnamefont {Aguilar}},\ and\ \bibinfo {author}
		{\bibfnamefont {A.~R.~H.}\ \bibnamefont {Walker}},\ }\bibfield  {title}
	{\bibinfo {title} {{Magnon-phonon hybridization in 2D antiferromagnet
				${\mathrm{MnPSe}}_{3}$}},\ }\href {https://doi.org/10.1126/sciadv.abj3106}
	{\bibfield  {journal} {\bibinfo  {journal} {Sci. Adv.}\ }\textbf {\bibinfo
			{volume} {7}},\ \bibinfo {pages} {eabj3106} (\bibinfo {year}
		{2021})}\BibitemShut {NoStop}%
	\bibitem [{\citenamefont {Squires}(1996)}]{squires1996introduction}%
	\BibitemOpen
	\bibfield  {author} {\bibinfo {author} {\bibfnamefont {G.~L.}\ \bibnamefont
			{Squires}},\ }\href@noop {} {\emph {\bibinfo {title} {{Introduction to the
					theory of thermal neutron scattering}}}}\ (\bibinfo  {publisher} {Courier
		Corporation},\ \bibinfo {year} {1996})\BibitemShut {NoStop}%
	\bibitem [{\citenamefont {Shirane}\ \emph {et~al.}(2002)\citenamefont
		{Shirane}, \citenamefont {Shapiro},\ and\ \citenamefont
		{Tranquada}}]{shirane2002neutron}%
	\BibitemOpen
	\bibfield  {author} {\bibinfo {author} {\bibfnamefont {G.}~\bibnamefont
			{Shirane}}, \bibinfo {author} {\bibfnamefont {S.~M.}\ \bibnamefont
			{Shapiro}},\ and\ \bibinfo {author} {\bibfnamefont {J.~M.}\ \bibnamefont
			{Tranquada}},\ }\href@noop {} {\emph {\bibinfo {title} {{Neutron scattering
					with a triple-axis spectrometer: basic techniques}}}}\ (\bibinfo  {publisher}
	{Cambridge University Press},\ \bibinfo {year} {2002})\BibitemShut {NoStop}%
	\bibitem [{\citenamefont {Du}\ \emph {et~al.}(2016)\citenamefont {Du},
		\citenamefont {Wang}, \citenamefont {Liu}, \citenamefont {Hu}, \citenamefont
		{Utama}, \citenamefont {Gan}, \citenamefont {Xiong},\ and\ \citenamefont
		{Kloc}}]{doi:10.1021/acsnano.5b05927}%
	\BibitemOpen
	\bibfield  {author} {\bibinfo {author} {\bibfnamefont {K.-Z.}\ \bibnamefont
			{Du}}, \bibinfo {author} {\bibfnamefont {X.-Z.}\ \bibnamefont {Wang}},
		\bibinfo {author} {\bibfnamefont {Y.}~\bibnamefont {Liu}}, \bibinfo {author}
		{\bibfnamefont {P.}~\bibnamefont {Hu}}, \bibinfo {author} {\bibfnamefont
			{M.~I.~B.}\ \bibnamefont {Utama}}, \bibinfo {author} {\bibfnamefont {C.~K.}\
			\bibnamefont {Gan}}, \bibinfo {author} {\bibfnamefont {Q.}~\bibnamefont
			{Xiong}},\ and\ \bibinfo {author} {\bibfnamefont {C.}~\bibnamefont {Kloc}},\
	}\bibfield  {title} {\bibinfo {title} {{Weak Van der Waals Stacking,
				Wide-Range Band Gap, and Raman Study on Ultrathin Layers of Metal Phosphorus
				Trichalcogenides}},\ }\href {https://doi.org/10.1021/acsnano.5b05927}
	{\bibfield  {journal} {\bibinfo  {journal} {ACS Nano}\ }\textbf {\bibinfo
			{volume} {10}},\ \bibinfo {pages} {1738} (\bibinfo {year}
		{2016})}\BibitemShut {NoStop}%
	\bibitem [{\citenamefont {Kajimoto}\ \emph {et~al.}(2011)\citenamefont
		{Kajimoto}, \citenamefont {Nakamura}, \citenamefont {Inamura}, \citenamefont
		{Mizuno}, \citenamefont {Nakajima}, \citenamefont {Ohira-Kawamura},
		\citenamefont {Yokoo}, \citenamefont {Nakatani}, \citenamefont {Maruyama},
		\citenamefont {Soyama}, \citenamefont {Shibata}, \citenamefont {Suzuya},
		\citenamefont {Sato}, \citenamefont {Aizawa}, \citenamefont {Arai},
		\citenamefont {Wakimoto}, \citenamefont {Ishikado}, \citenamefont {Shamoto},
		\citenamefont {Fujita}, \citenamefont {Hiraka}, \citenamefont {Ohoyama},
		\citenamefont {Yamada},\ and\ \citenamefont {Lee}}]{Kajimo}%
	\BibitemOpen
	\bibfield  {author} {\bibinfo {author} {\bibfnamefont {R.}~\bibnamefont
			{Kajimoto}}, \bibinfo {author} {\bibfnamefont {M.}~\bibnamefont {Nakamura}},
		\bibinfo {author} {\bibfnamefont {Y.}~\bibnamefont {Inamura}}, \bibinfo
		{author} {\bibfnamefont {F.}~\bibnamefont {Mizuno}}, \bibinfo {author}
		{\bibfnamefont {K.}~\bibnamefont {Nakajima}}, \bibinfo {author}
		{\bibfnamefont {S.}~\bibnamefont {Ohira-Kawamura}}, \bibinfo {author}
		{\bibfnamefont {T.}~\bibnamefont {Yokoo}}, \bibinfo {author} {\bibfnamefont
			{T.}~\bibnamefont {Nakatani}}, \bibinfo {author} {\bibfnamefont
			{R.}~\bibnamefont {Maruyama}}, \bibinfo {author} {\bibfnamefont
			{K.}~\bibnamefont {Soyama}}, \bibinfo {author} {\bibfnamefont
			{K.}~\bibnamefont {Shibata}}, \bibinfo {author} {\bibfnamefont
			{K.}~\bibnamefont {Suzuya}}, \bibinfo {author} {\bibfnamefont
			{S.}~\bibnamefont {Sato}}, \bibinfo {author} {\bibfnamefont {K.}~\bibnamefont
			{Aizawa}}, \bibinfo {author} {\bibfnamefont {M.}~\bibnamefont {Arai}},
		\bibinfo {author} {\bibfnamefont {S.}~\bibnamefont {Wakimoto}}, \bibinfo
		{author} {\bibfnamefont {M.}~\bibnamefont {Ishikado}}, \bibinfo {author}
		{\bibfnamefont {S.}~\bibnamefont {Shamoto}}, \bibinfo {author} {\bibfnamefont
			{M.}~\bibnamefont {Fujita}}, \bibinfo {author} {\bibfnamefont
			{H.}~\bibnamefont {Hiraka}}, \bibinfo {author} {\bibfnamefont
			{K.}~\bibnamefont {Ohoyama}}, \bibinfo {author} {\bibfnamefont
			{K.}~\bibnamefont {Yamada}},\ and\ \bibinfo {author} {\bibfnamefont
			{C.}~\bibnamefont {Lee}},\ }\bibfield  {title} {\bibinfo {title} {{The Fermi
				Chopper Spectrometer 4SEASONS at J-PARC}},\ }\href
	{https://doi.org/10.1143/JPSJS.80SB.SB025} {\bibfield  {journal} {\bibinfo
			{journal} {J. Phys. Soc. Jpn}\ }\textbf {\bibinfo {volume} {80}},\ \bibinfo
		{pages} {SB025} (\bibinfo {year} {2011})}\BibitemShut {NoStop}%
	\bibitem [{\citenamefont {Nakamura}\ \emph {et~al.}(2009)\citenamefont
		{Nakamura}, \citenamefont {Kajimoto}, \citenamefont {Inamura}, \citenamefont
		{Mizuno}, \citenamefont {Fujita}, \citenamefont {Yokoo},\ and\ \citenamefont
		{Arai}}]{doi:10.1143/JPSJ.78.093002}%
	\BibitemOpen
	\bibfield  {author} {\bibinfo {author} {\bibfnamefont {M.}~\bibnamefont
			{Nakamura}}, \bibinfo {author} {\bibfnamefont {R.}~\bibnamefont {Kajimoto}},
		\bibinfo {author} {\bibfnamefont {Y.}~\bibnamefont {Inamura}}, \bibinfo
		{author} {\bibfnamefont {F.}~\bibnamefont {Mizuno}}, \bibinfo {author}
		{\bibfnamefont {M.}~\bibnamefont {Fujita}}, \bibinfo {author} {\bibfnamefont
			{T.}~\bibnamefont {Yokoo}},\ and\ \bibinfo {author} {\bibfnamefont
			{M.}~\bibnamefont {Arai}},\ }\bibfield  {title} {\bibinfo {title} {{First
				Demonstration of Novel Method for Inelastic Neutron Scattering Measurement
				Utilizing Multiple Incident Energies}},\ }\href
	{https://doi.org/10.1143/JPSJ.78.093002} {\bibfield  {journal} {\bibinfo
			{journal} {J. Phys. Soc. Jpn}\ }\textbf {\bibinfo {volume} {78}},\ \bibinfo
		{pages} {093002} (\bibinfo {year} {2009})}\BibitemShut {NoStop}%
	\bibitem [{\citenamefont {Inamura}\ \emph {et~al.}(2013)\citenamefont
		{Inamura}, \citenamefont {Nakatani}, \citenamefont {Suzuki},\ and\
		\citenamefont {Otomo}}]{doi:10.7566/JPSJS.82SA.SA031}%
	\BibitemOpen
	\bibfield  {author} {\bibinfo {author} {\bibfnamefont {Y.}~\bibnamefont
			{Inamura}}, \bibinfo {author} {\bibfnamefont {T.}~\bibnamefont {Nakatani}},
		\bibinfo {author} {\bibfnamefont {J.}~\bibnamefont {Suzuki}},\ and\ \bibinfo
		{author} {\bibfnamefont {T.}~\bibnamefont {Otomo}},\ }\bibfield  {title}
	{\bibinfo {title} {{Development Status of Software “Utsusemi” for Chopper
				Spectrometers at MLF, J-PARC}},\ }\href
	{https://doi.org/10.7566/JPSJS.82SA.SA031} {\bibfield  {journal} {\bibinfo
			{journal} {J. Phys. Soc. Jpn}\ }\textbf {\bibinfo {volume} {82}},\ \bibinfo
		{pages} {SA031} (\bibinfo {year} {2013})}\BibitemShut {NoStop}%
	\bibitem [{\citenamefont {Ewings}\ \emph {et~al.}(2016)\citenamefont {Ewings},
		\citenamefont {Buts}, \citenamefont {Le}, \citenamefont {{van Duijn}},
		\citenamefont {Bustinduy},\ and\ \citenamefont {Perring}}]{EWINGS2016132}%
	\BibitemOpen
	\bibfield  {author} {\bibinfo {author} {\bibfnamefont {R.~A.}\ \bibnamefont
			{Ewings}}, \bibinfo {author} {\bibfnamefont {A.}~\bibnamefont {Buts}},
		\bibinfo {author} {\bibfnamefont {M.}~\bibnamefont {Le}}, \bibinfo {author}
		{\bibfnamefont {J.}~\bibnamefont {{van Duijn}}}, \bibinfo {author}
		{\bibfnamefont {I.}~\bibnamefont {Bustinduy}},\ and\ \bibinfo {author}
		{\bibfnamefont {T.}~\bibnamefont {Perring}},\ }\bibfield  {title} {\bibinfo
		{title} {{Horace: Software for the analysis of data from single crystal
				spectroscopy experiments at time-of-flight neutron instruments}},\ }\href
	{https://doi.org/https://doi.org/10.1016/j.nima.2016.07.036} {\bibfield
		{journal} {\bibinfo  {journal} {Nucl. Instrum. Methods Phys. Res., Sect. A}\
		}\textbf {\bibinfo {volume} {834}},\ \bibinfo {pages} {132} (\bibinfo {year}
		{2016})}\BibitemShut {NoStop}%
	\bibitem [{\citenamefont {Toth}\ and\ \citenamefont {Lake}(2015)}]{Toth_2015}%
	\BibitemOpen
	\bibfield  {author} {\bibinfo {author} {\bibfnamefont {S.}~\bibnamefont
			{Toth}}\ and\ \bibinfo {author} {\bibfnamefont {B.}~\bibnamefont {Lake}},\
	}\bibfield  {title} {\bibinfo {title} {{Linear spin wave theory for single-Q
				incommensurate magnetic structures}},\ }\href
	{https://doi.org/10.1088/0953-8984/27/16/166002} {\bibfield  {journal}
		{\bibinfo  {journal} {J. Phys. Condens. Matter}\ }\textbf {\bibinfo {volume}
			{27}},\ \bibinfo {pages} {166002} (\bibinfo {year} {2015})}\BibitemShut
	{NoStop}%
	\bibitem [{\citenamefont {Jana}\ \emph {et~al.}(2023)\citenamefont {Jana},
		\citenamefont {Vaclavkova}, \citenamefont {Mohelsky}, \citenamefont
		{Kapuscinski}, \citenamefont {Cho}, \citenamefont {Breslavetz}, \citenamefont
		{Bia艂ek}, \citenamefont {Ansermet}, \citenamefont {Piot}, \citenamefont
		{Orlita}, \citenamefont {Faugeras},\ and\ \citenamefont
		{Potemski}}]{jana2023magnon}%
	\BibitemOpen
	\bibfield  {author} {\bibinfo {author} {\bibfnamefont {D.}~\bibnamefont
			{Jana}}, \bibinfo {author} {\bibfnamefont {D.}~\bibnamefont {Vaclavkova}},
		\bibinfo {author} {\bibfnamefont {I.}~\bibnamefont {Mohelsky}}, \bibinfo
		{author} {\bibfnamefont {P.}~\bibnamefont {Kapuscinski}}, \bibinfo {author}
		{\bibfnamefont {C.~W.}\ \bibnamefont {Cho}}, \bibinfo {author} {\bibfnamefont
			{I.}~\bibnamefont {Breslavetz}}, \bibinfo {author} {\bibfnamefont
			{M.}~\bibnamefont {Białek}}, \bibinfo {author} {\bibfnamefont {J.~P.}\
			\bibnamefont {Ansermet}}, \bibinfo {author} {\bibfnamefont {B.~A.}\
			\bibnamefont {Piot}}, \bibinfo {author} {\bibfnamefont {M.}~\bibnamefont
			{Orlita}}, \bibinfo {author} {\bibfnamefont {C.}~\bibnamefont {Faugeras}},\
		and\ \bibinfo {author} {\bibfnamefont {M.}~\bibnamefont {Potemski}},\
	}\href@noop {} {\bibinfo {title} {{Magnon gap excitations in van der Waals
			antiferromagnet MnPSe$_3$}}} (\bibinfo {year} {2023}),\ \Eprint
			{https://arxiv.org/abs/2309.06866} {arXiv:2309.06866 [cond-mat.mtrl-sci]}
	\BibitemShut {NoStop}%
	\bibitem [{\citenamefont {Nolting}\ and\ \citenamefont
		{Ramakanth}(2009)}]{nolting2009quantum}%
	\BibitemOpen
	\bibfield  {author} {\bibinfo {author} {\bibfnamefont {W.}~\bibnamefont
			{Nolting}}\ and\ \bibinfo {author} {\bibfnamefont {A.}~\bibnamefont
			{Ramakanth}},\ }\href@noop {} {\emph {\bibinfo {title} {Quantum theory of
				magnetism}}}\ (\bibinfo  {publisher} {Springer Science \& Business Media},\
	\bibinfo {year} {2009})\BibitemShut {NoStop}%
	\bibitem [{\citenamefont {Takahashi}\ and\ \citenamefont
		{Nagaosa}(2016)}]{PhysRevLett.117.217205}%
	\BibitemOpen
	\bibfield  {author} {\bibinfo {author} {\bibfnamefont {R.}~\bibnamefont
			{Takahashi}}\ and\ \bibinfo {author} {\bibfnamefont {N.}~\bibnamefont
			{Nagaosa}},\ }\bibfield  {title} {\bibinfo {title} {{Berry Curvature in
				Magnon-Phonon Hybrid Systems}},\ }\href
	{https://doi.org/10.1103/PhysRevLett.117.217205} {\bibfield  {journal}
		{\bibinfo  {journal} {Phys. Rev. Lett.}\ }\textbf {\bibinfo {volume} {117}},\
		\bibinfo {pages} {217205} (\bibinfo {year} {2016})}\BibitemShut {NoStop}%
	\bibitem [{\citenamefont {Zhang}\ \emph {et~al.}(2019)\citenamefont {Zhang},
		\citenamefont {Zhang}, \citenamefont {Okamoto},\ and\ \citenamefont
		{Xiao}}]{PhysRevLett.123.167202}%
	\BibitemOpen
	\bibfield  {author} {\bibinfo {author} {\bibfnamefont {X.}~\bibnamefont
			{Zhang}}, \bibinfo {author} {\bibfnamefont {Y.}~\bibnamefont {Zhang}},
		\bibinfo {author} {\bibfnamefont {S.}~\bibnamefont {Okamoto}},\ and\ \bibinfo
		{author} {\bibfnamefont {D.}~\bibnamefont {Xiao}},\ }\bibfield  {title}
	{\bibinfo {title} {{Thermal Hall Effect Induced by Magnon-Phonon
				Interactions}},\ }\href {https://doi.org/10.1103/PhysRevLett.123.167202}
	{\bibfield  {journal} {\bibinfo  {journal} {Phys. Rev. Lett.}\ }\textbf
		{\bibinfo {volume} {123}},\ \bibinfo {pages} {167202} (\bibinfo {year}
		{2019})}\BibitemShut {NoStop}%
	\bibitem [{\citenamefont {Kikkawa}\ \emph {et~al.}(2016)\citenamefont
		{Kikkawa}, \citenamefont {Shen}, \citenamefont {Flebus}, \citenamefont
		{Duine}, \citenamefont {Uchida}, \citenamefont {Qiu}, \citenamefont {Bauer},\
		and\ \citenamefont {Saitoh}}]{PhysRevLett.117.207203}%
	\BibitemOpen
	\bibfield  {author} {\bibinfo {author} {\bibfnamefont {T.}~\bibnamefont
			{Kikkawa}}, \bibinfo {author} {\bibfnamefont {K.}~\bibnamefont {Shen}},
		\bibinfo {author} {\bibfnamefont {B.}~\bibnamefont {Flebus}}, \bibinfo
		{author} {\bibfnamefont {R.~A.}\ \bibnamefont {Duine}}, \bibinfo {author}
		{\bibfnamefont {K.}~\bibnamefont {Uchida}}, \bibinfo {author} {\bibfnamefont
			{Z.}~\bibnamefont {Qiu}}, \bibinfo {author} {\bibfnamefont {G.~E.~W.}\
			\bibnamefont {Bauer}},\ and\ \bibinfo {author} {\bibfnamefont
			{E.}~\bibnamefont {Saitoh}},\ }\bibfield  {title} {\bibinfo {title} {{Magnon
				Polarons in the Spin Seebeck Effect}},\ }\href
	{https://doi.org/10.1103/PhysRevLett.117.207203} {\bibfield  {journal}
		{\bibinfo  {journal} {Phys. Rev. Lett.}\ }\textbf {\bibinfo {volume} {117}},\
		\bibinfo {pages} {207203} (\bibinfo {year} {2016})}\BibitemShut {NoStop}%
	\bibitem [{\citenamefont {Bao}\ \emph {et~al.}(2023)\citenamefont {Bao},
		\citenamefont {Gu}, \citenamefont {Shangguan}, \citenamefont {Huang},
		\citenamefont {Liao}, \citenamefont {Zhao}, \citenamefont {Zhang},
		\citenamefont {Dong}, \citenamefont {Wang}, \citenamefont {Kajimoto} \emph
		{et~al.}}]{NC14_6093}%
	\BibitemOpen
	\bibfield  {author} {\bibinfo {author} {\bibfnamefont {S.}~\bibnamefont
			{Bao}}, \bibinfo {author} {\bibfnamefont {Z.-L.}\ \bibnamefont {Gu}},
		\bibinfo {author} {\bibfnamefont {Y.}~\bibnamefont {Shangguan}}, \bibinfo
		{author} {\bibfnamefont {Z.}~\bibnamefont {Huang}}, \bibinfo {author}
		{\bibfnamefont {J.}~\bibnamefont {Liao}}, \bibinfo {author} {\bibfnamefont
			{X.}~\bibnamefont {Zhao}}, \bibinfo {author} {\bibfnamefont {B.}~\bibnamefont
			{Zhang}}, \bibinfo {author} {\bibfnamefont {Z.-Y.}\ \bibnamefont {Dong}},
		\bibinfo {author} {\bibfnamefont {W.}~\bibnamefont {Wang}}, \bibinfo {author}
		{\bibfnamefont {R.}~\bibnamefont {Kajimoto}}, \emph {et~al.},\ }\bibfield
	{title} {\bibinfo {title} {{Direct observation of topological magnon polarons
				in a multiferroic material}},\ }\href
	{https://doi.org/https://doi.org/10.1038/s41467-023-41791-9} {\bibfield
		{journal} {\bibinfo  {journal} {Nat. Commun.}\ }\textbf {\bibinfo {volume}
			{14}},\ \bibinfo {pages} {6093} (\bibinfo {year} {2023})}\BibitemShut
	{NoStop}%
	\bibitem [{\citenamefont {Go}\ \emph {et~al.}(2019)\citenamefont {Go},
		\citenamefont {Kim},\ and\ \citenamefont {Lee}}]{PhysRevLett.123.237207}%
	\BibitemOpen
	\bibfield  {author} {\bibinfo {author} {\bibfnamefont {G.}~\bibnamefont
			{Go}}, \bibinfo {author} {\bibfnamefont {S.~K.}\ \bibnamefont {Kim}},\ and\
		\bibinfo {author} {\bibfnamefont {K.}~\bibnamefont {Lee}},\ }\bibfield
	{title} {\bibinfo {title} {{Topological Magnon-Phonon Hybrid Excitations in
				Two-Dimensional Ferromagnets with Tunable Chern Numbers}},\ }\href
	{https://doi.org/10.1103/PhysRevLett.123.237207} {\bibfield  {journal}
		{\bibinfo  {journal} {Phys. Rev. Lett.}\ }\textbf {\bibinfo {volume} {123}},\
		\bibinfo {pages} {237207} (\bibinfo {year} {2019})}\BibitemShut {NoStop}%
	\bibitem [{\citenamefont {Shen}\ and\ \citenamefont
		{Kim}(2020)}]{PhysRevB.101.125111}%
	\BibitemOpen
	\bibfield  {author} {\bibinfo {author} {\bibfnamefont {P.}~\bibnamefont
			{Shen}}\ and\ \bibinfo {author} {\bibfnamefont {S.~K.}\ \bibnamefont {Kim}},\
	}\bibfield  {title} {\bibinfo {title} {{Magnetic field control of topological
				magnon-polaron bands in two-dimensional ferromagnets}},\ }\href
	{https://doi.org/10.1103/PhysRevB.101.125111} {\bibfield  {journal} {\bibinfo
			{journal} {Phys. Rev. B}\ }\textbf {\bibinfo {volume} {101}},\ \bibinfo
		{pages} {125111} (\bibinfo {year} {2020})}\BibitemShut {NoStop}%
	\bibitem [{\citenamefont {Zhang}\ \emph {et~al.}(2020)\citenamefont {Zhang},
		\citenamefont {Go}, \citenamefont {Lee},\ and\ \citenamefont
		{Kim}}]{PhysRevLett.124.147204}%
	\BibitemOpen
	\bibfield  {author} {\bibinfo {author} {\bibfnamefont {S.}~\bibnamefont
			{Zhang}}, \bibinfo {author} {\bibfnamefont {G.}~\bibnamefont {Go}}, \bibinfo
		{author} {\bibfnamefont {K.}~\bibnamefont {Lee}},\ and\ \bibinfo {author}
		{\bibfnamefont {S.~K.}\ \bibnamefont {Kim}},\ }\bibfield  {title} {\bibinfo
		{title} {{SU(3) Topology of Magnon-Phonon Hybridization in 2D
				Antiferromagnets}},\ }\href {https://doi.org/10.1103/PhysRevLett.124.147204}
	{\bibfield  {journal} {\bibinfo  {journal} {Phys. Rev. Lett.}\ }\textbf
		{\bibinfo {volume} {124}},\ \bibinfo {pages} {147204} (\bibinfo {year}
		{2020})}\BibitemShut {NoStop}%
	\bibitem [{\citenamefont {Kittel}(1949)}]{RevModPhys.21.541}%
	\BibitemOpen
	\bibfield  {author} {\bibinfo {author} {\bibfnamefont {C.}~\bibnamefont
			{Kittel}},\ }\bibfield  {title} {\bibinfo {title} {{Physical Theory of
				Ferromagnetic Domains}},\ }\href {https://doi.org/10.1103/RevModPhys.21.541}
	{\bibfield  {journal} {\bibinfo  {journal} {Rev. Mod. Phys.}\ }\textbf
		{\bibinfo {volume} {21}},\ \bibinfo {pages} {541} (\bibinfo {year}
		{1949})}\BibitemShut {NoStop}%
	\bibitem [{\citenamefont {Kittel}(1958)}]{PhysRev.110.836}%
	\BibitemOpen
	\bibfield  {author} {\bibinfo {author} {\bibfnamefont {C.}~\bibnamefont
			{Kittel}},\ }\bibfield  {title} {\bibinfo {title} {{Interaction of Spin Waves
				and Ultrasonic Waves in Ferromagnetic Crystals}},\ }\href
	{https://doi.org/10.1103/PhysRev.110.836} {\bibfield  {journal} {\bibinfo
			{journal} {Phys. Rev.}\ }\textbf {\bibinfo {volume} {110}},\ \bibinfo {pages}
		{836} (\bibinfo {year} {1958})}\BibitemShut {NoStop}%
	\bibitem [{\citenamefont {Ma}\ and\ \citenamefont
		{Fiete}(2022)}]{PhysRevB.105.L100402}%
	\BibitemOpen
	\bibfield  {author} {\bibinfo {author} {\bibfnamefont {B.}~\bibnamefont
			{Ma}}\ and\ \bibinfo {author} {\bibfnamefont {G.~A.}\ \bibnamefont {Fiete}},\
	}\bibfield  {title} {\bibinfo {title} {{Antiferromagnetic insulators with
				tunable magnon-polaron Chern numbers induced by in-plane optical phonons}},\
	}\href {https://doi.org/10.1103/PhysRevB.105.L100402} {\bibfield  {journal}
		{\bibinfo  {journal} {Phys. Rev. B}\ }\textbf {\bibinfo {volume} {105}},\
		\bibinfo {pages} {L100402} (\bibinfo {year} {2022})}\BibitemShut {NoStop}%
	\bibitem [{\citenamefont {Park}\ \emph {et~al.}(2020)\citenamefont {Park},
		\citenamefont {Nagaosa},\ and\ \citenamefont {Yang}}]{ParKS}%
	\BibitemOpen
	\bibfield  {author} {\bibinfo {author} {\bibfnamefont {S.}~\bibnamefont
			{Park}}, \bibinfo {author} {\bibfnamefont {N.}~\bibnamefont {Nagaosa}},\ and\
		\bibinfo {author} {\bibfnamefont {B.}~\bibnamefont {Yang}},\ }\bibfield
	{title} {\bibinfo {title} {{Thermal Hall Effect, Spin Nernst Effect, and Spin
				Density Induced by a Thermal Gradient in Collinear Ferrimagnets from
				Magnon-phonon Interaction}},\ }\href
	{https://doi.org/10.1021/acs.nanolett.0c00363} {\bibfield  {journal}
		{\bibinfo  {journal} {Nano Lett.}\ }\textbf {\bibinfo {volume} {20}},\
		\bibinfo {pages} {2741} (\bibinfo {year} {2020})}\BibitemShut {NoStop}%
	\bibitem [{\citenamefont {Johnston}(2016)}]{PhysRevB.93.014421}%
	\BibitemOpen
	\bibfield  {author} {\bibinfo {author} {\bibfnamefont {D.~C.}\ \bibnamefont
			{Johnston}},\ }\bibfield  {title} {\bibinfo {title} {{Magnetic dipole
				interactions in crystals}},\ }\href
	{https://doi.org/10.1103/PhysRevB.93.014421} {\bibfield  {journal} {\bibinfo
			{journal} {Phys. Rev. B}\ }\textbf {\bibinfo {volume} {93}},\ \bibinfo
		{pages} {014421} (\bibinfo {year} {2016})}\BibitemShut {NoStop}%
	\bibitem [{\citenamefont {W.~Baltensperger}(1968)}]{Baltensperger}%
	\BibitemOpen
	\bibfield  {author} {\bibinfo {author} {\bibfnamefont {J.~S.~H.}\
			\bibnamefont {W.~Baltensperger}},\ }\bibfield  {title} {\bibinfo {title}
		{{Influence of magnetic order in insulators on the optical phonon
				frequency}},\ }\href
	{https://www.e-periodica.ch/cntmng?pid=hpa-001%3A1968%3A41%3A%3A1447}
	{\bibfield  {journal} {\bibinfo  {journal} {Helv. Phys. Acta}\ }\textbf
		{\bibinfo {volume} {41}},\ \bibinfo {pages} {668} (\bibinfo {year}
		{1968})}\BibitemShut {NoStop}%
	\bibitem [{\citenamefont {Huberman}\ \emph {et~al.}(2005)\citenamefont
		{Huberman}, \citenamefont {Coldea}, \citenamefont {Cowley}, \citenamefont
		{Tennant}, \citenamefont {Leheny}, \citenamefont {Christianson},\ and\
		\citenamefont {Frost}}]{PhysRevB.72.014413}%
	\BibitemOpen
	\bibfield  {author} {\bibinfo {author} {\bibfnamefont {T.}~\bibnamefont
			{Huberman}}, \bibinfo {author} {\bibfnamefont {R.}~\bibnamefont {Coldea}},
		\bibinfo {author} {\bibfnamefont {R.~A.}\ \bibnamefont {Cowley}}, \bibinfo
		{author} {\bibfnamefont {D.~A.}\ \bibnamefont {Tennant}}, \bibinfo {author}
		{\bibfnamefont {R.~L.}\ \bibnamefont {Leheny}}, \bibinfo {author}
		{\bibfnamefont {R.~J.}\ \bibnamefont {Christianson}},\ and\ \bibinfo {author}
		{\bibfnamefont {C.~D.}\ \bibnamefont {Frost}},\ }\bibfield  {title} {\bibinfo
		{title} {{Two-magnon excitations observed by neutron scattering in the
				two-dimensional spin-$\frac{5}{2}$ Heisenberg antiferromagnet
				${\mathrm{Rb}}_{2}\mathrm{Mn}{\mathrm{F}}_{4}$}},\ }\href
	{https://doi.org/10.1103/PhysRevB.72.014413} {\bibfield  {journal} {\bibinfo
			{journal} {Phys. Rev. B}\ }\textbf {\bibinfo {volume} {72}},\ \bibinfo
		{pages} {014413} (\bibinfo {year} {2005})}\BibitemShut {NoStop}%
	\bibitem [{\citenamefont {Cowley}\ \emph {et~al.}(1969)\citenamefont {Cowley},
		\citenamefont {Buyers}, \citenamefont {Martel},\ and\ \citenamefont
		{Stevenson}}]{PhysRevLett.23.86}%
	\BibitemOpen
	\bibfield  {author} {\bibinfo {author} {\bibfnamefont {R.~A.}\ \bibnamefont
			{Cowley}}, \bibinfo {author} {\bibfnamefont {W.~J.~L.}\ \bibnamefont
			{Buyers}}, \bibinfo {author} {\bibfnamefont {P.}~\bibnamefont {Martel}},\
		and\ \bibinfo {author} {\bibfnamefont {R.~W.~H.}\ \bibnamefont {Stevenson}},\
	}\bibfield  {title} {\bibinfo {title} {{Two-Magnon Scattering of Neutrons}},\
	}\href {https://doi.org/10.1103/PhysRevLett.23.86} {\bibfield  {journal}
		{\bibinfo  {journal} {Phys. Rev. Lett.}\ }\textbf {\bibinfo {volume} {23}},\
		\bibinfo {pages} {86} (\bibinfo {year} {1969})}\BibitemShut {NoStop}%
	\bibitem [{\citenamefont {Park}\ \emph {et~al.}(2016)\citenamefont {Park},
		\citenamefont {Oh}, \citenamefont {Leiner}, \citenamefont {Jeong},
		\citenamefont {Rule}, \citenamefont {Le},\ and\ \citenamefont
		{Park}}]{PhysRevB.94.104421}%
	\BibitemOpen
	\bibfield  {author} {\bibinfo {author} {\bibfnamefont {K.}~\bibnamefont
			{Park}}, \bibinfo {author} {\bibfnamefont {J.}~\bibnamefont {Oh}}, \bibinfo
		{author} {\bibfnamefont {J.~C.}\ \bibnamefont {Leiner}}, \bibinfo {author}
		{\bibfnamefont {J.}~\bibnamefont {Jeong}}, \bibinfo {author} {\bibfnamefont
			{K.~C.}\ \bibnamefont {Rule}}, \bibinfo {author} {\bibfnamefont {M.~D.}\
			\bibnamefont {Le}},\ and\ \bibinfo {author} {\bibfnamefont {J.}~\bibnamefont
			{Park}},\ }\bibfield  {title} {\bibinfo {title} {{Magnon-phonon coupling and
				two-magnon continuum in the two-dimensional triangular antiferromagnet
				${\mathrm{CuCrO}}_{2}$}},\ }\href
	{https://doi.org/10.1103/PhysRevB.94.104421} {\bibfield  {journal} {\bibinfo
			{journal} {Phys. Rev. B}\ }\textbf {\bibinfo {volume} {94}},\ \bibinfo
		{pages} {104421} (\bibinfo {year} {2016})}\BibitemShut {NoStop}%
	\bibitem [{\citenamefont {Songvilay}\ \emph {et~al.}(2021)\citenamefont
		{Songvilay}, \citenamefont {Petit}, \citenamefont {Damay}, \citenamefont
		{Roux}, \citenamefont {Qureshi}, \citenamefont {Walker}, \citenamefont
		{Rodriguez-Rivera}, \citenamefont {Gao}, \citenamefont {Cheong},\ and\
		\citenamefont {Stock}}]{PhysRevLett.126.017201}%
	\BibitemOpen
	\bibfield  {author} {\bibinfo {author} {\bibfnamefont {M.}~\bibnamefont
			{Songvilay}}, \bibinfo {author} {\bibfnamefont {S.}~\bibnamefont {Petit}},
		\bibinfo {author} {\bibfnamefont {F.}~\bibnamefont {Damay}}, \bibinfo
		{author} {\bibfnamefont {G.}~\bibnamefont {Roux}}, \bibinfo {author}
		{\bibfnamefont {N.}~\bibnamefont {Qureshi}}, \bibinfo {author} {\bibfnamefont
			{H.~C.}\ \bibnamefont {Walker}}, \bibinfo {author} {\bibfnamefont {J.~A.}\
			\bibnamefont {Rodriguez-Rivera}}, \bibinfo {author} {\bibfnamefont
			{B.}~\bibnamefont {Gao}}, \bibinfo {author} {\bibfnamefont {S.-W.}\
			\bibnamefont {Cheong}},\ and\ \bibinfo {author} {\bibfnamefont
			{C.}~\bibnamefont {Stock}},\ }\bibfield  {title} {\bibinfo {title} {{From
				One- to Two-Magnon Excitations in the $S=3/2$ Magnet
				$\ensuremath{\beta}\text{\ensuremath{-}}{\mathrm{CaCr}}_{2}{\mathrm{O}}_{4}$}},\
	}\href {https://doi.org/10.1103/PhysRevLett.126.017201} {\bibfield  {journal}
		{\bibinfo  {journal} {Phys. Rev. Lett.}\ }\textbf {\bibinfo {volume} {126}},\
		\bibinfo {pages} {017201} (\bibinfo {year} {2021})}\BibitemShut {NoStop}%
\end{thebibliography}
%

\end{document}